\documentclass[aps,prb,reprint,superscriptaddress]{revtex4-1}

\usepackage{graphicx}
\usepackage{hyperref}

\begin{document}


\title{Non-adiabatic effects and exciton-like states during insulator-to-metal transition in warm dense hydrogen}


\author{Ilya~D.~Fedorov}
    \email[]{fedorov.id@mipt.ru}
    \affiliation{Joint Institute for High Temperatures of the Russian Academy of Sciences (JIHT RAS), Izhorskaya 13 Building 2, Moscow 125412, Russian Federation}
    \affiliation{Moscow Institute of Physics and Technologies National Research University (MIPT NRU), Institutskij pereulok 9, Dolgoprudny Moscow region 141700, Russian Federation}
\author{Nikita~D.~Orekhov}
    \affiliation{Joint Institute for High Temperatures of the Russian Academy of Sciences (JIHT RAS), Izhorskaya 13 Building 2, Moscow 125412, Russian Federation}
    \affiliation{Moscow Institute of Physics and Technologies National Research University (MIPT NRU), Institutskij pereulok 9, Dolgoprudny Moscow region 141700, Russian Federation}
\author{Vladimir~V.~Stegailov}
    \email[]{stegailov.vv@mipt.ru}
    \affiliation{Joint Institute for High Temperatures of the Russian Academy of Sciences (JIHT RAS), Izhorskaya 13 Building 2, Moscow 125412, Russian Federation}
    \affiliation{Moscow Institute of Physics and Technologies National Research University (MIPT NRU), Institutskij pereulok 9, Dolgoprudny Moscow region 141700, Russian Federation}
    \affiliation{National Research University Higher School of Economics (NRU HSE), Myasnitskaya ulitsa 20, Moscow 101000 Russian Federation}


\date{\today}

\begin{abstract}
Transition of molecular hydrogen to atomic ionized state with increase of temperature and pressure poses still unresolved problems for experimental methods and theory. Here we analyze the dynamics of this transition and show its nonequilibrium non-adiabatic character overlooked in both interpreting experimental data and in theoretical models. The non-adiabatic mechanism explains the strong isotopic effect \href{https://doi.org/10.1103/PhysRevB.98.104102}{[Zaghoo, Husband, and Silvera, Phys. Rev. B 98, 104102 (2018)]} and the large latent heat \href{https://doi.org/10.1103/PhysRevB.98.104102}{[Houtput, Tempere, and Silvera, Phys. Rev. B 100, 134106 (2019)]} reported recently. We demonstrate the possibility of formation of intermediate exciton-like molecular states at heating of molecular hydrogen that can explain puzzling experimental data on reflectivity and conductivity during the insulator-to-metal transition.
\end{abstract}


\maketitle

\textit{Introduction.} The nature of the molecular-to-atomic transition in fluid hydrogen and deuterium is a fundamental problem~\cite{mcmahon2012properties,Utyuzh_2017} that has been drawing increasing attention for more than two decades since the first reliable experiments on electrical conductivities of fluid H$_2$/D$_2$ at shock pressures~\cite{Weir-etal-PhysRevLett.76.1860-1996}. In subsequent dynamic and static experiments a large amount of experimental data has been collected (e.g., see~\cite{Fortov-etal-PhysRevLett.99.185001-2007,Knudson1455,Mochalov2017,Celliers2018} and~\cite{Loubeyre-etal-HighPressRes-2004,Dzyabura2013,ohta2015phase,goncharov2016,ZaghooHusbandSilvera-PhysRevB.98.104102-2018}). However at the moment, there are inconsistencies between different experimental results and there is no complete theoretical understanding of this transition. 

The first theoretical approaches to the equation of state of warm dense hydrogen were based on chemical models~\cite{norman1968insufficiency,Biberman1969,Lebowitz1969,Norman1970,Norman1970a,Ebeling1971,Kraeft1986,Saumon1991,Reinholz1995,Ebeling2003,Norman2006,Khomkin2013,Starostin2016,Ebeling2017}. The possibility of plasma phase transition was predicted~\cite{norman1968insufficiency,FilinovNorman-PRA-1975}. Later, the concept of insulator-to-metal transition (IMT) became widespread for interpreting results of \textit{ab initio} calculations~\cite{Redmer2010book}. The first-principles molecular dynamics (FPMD) based on density functional theory (DFT) and the quantum Monte-Carlo (QMC) methods are considered as the most accurate theoretical tools for calculations of IMT in hydrogen (e.g. see~\cite{Scandolo-PNAS2003,TamblynBonev-PhysRevLett.104.065702-2010,Morales-et-al-PhysRevLett.110.065702-2013,Mazolla2017,Knudson-Desjarlais-PhysRevLett.118.035501-2017,Knudson-etal-PhysRevB.98.174110-2018,Ackland-etal-PhysRevB.100.134109-2019} and~\cite{DelaneyPierleoniCeperley-PhysRevLett.97.235702-2006,Tubman-etal-PhysRevLett.115.045301-2015,MazzolaSorella-PhysRevLett.114.105701-2015,Mazzola-etal-PhysRevLett.120.025701-2018,Rillo9770}). These methods have been systematically developed and are able to take into account the coupling of electrons and nuclei in QMC~\cite{DelaneyPierleoniCeperley-PhysRevLett.97.235702-2006}, the nuclear quantum effects (NQE)~\cite{Morales-et-al-PhysRevLett.110.065702-2013}, dispersion interactions~\cite{Morales-et-al-PhysRevLett.110.065702-2013,Mazolla2017}, the influence of a particular choice of xc-functional in DFT~\cite{Knudson-Desjarlais-PhysRevLett.118.035501-2017,Knudson-etal-PhysRevB.98.174110-2018} and xc-functional dependence on electronic temperature~\cite{KarasievDuftyTrickey-PhysRevLett.120.076401-2018}.

Two important common features for all these methods are the assumption of the thermodynamic equilibrium of the nuclear and electronic subsystems and the adiabatic Born-Oppenheimer approximation. In QMC the ground state of the many-electron system is calculated for independent nuclear configurations sampling canonical ensemble~\cite{DelaneyPierleoniCeperley-PhysRevLett.97.235702-2006}. In FPMD DFT methods the Fermi-Dirac distribution is assumed for Kohn-Sham electron states that corresponds to the Mermin finite-temperature DFT (FT-DFT) formalism~\cite{Mermin-PhysRev.137.A1441-1965}. Correspondingly, the experiments on IMT are interpreted within the thermodynamic equilibrium framework. However, as far as we are aware of, there has been no careful analysis of the possible non-equilibrium effects on IMT in fluid H$_2$/D$_2$. Low electron-ion recombination and temperature relaxation rates in warm dense hydrogen~\cite{LankinNorman-CPP-2009,eff2019-Murillo} have pointed to the importance of the non-equilibrium effects but in the atomic state of fluid H$_2$/D$_2$ at temperatures higher than IMT.

In this Rapid Communication we would like to put the focus on the non-equilibrium non-adiabatic processes accompanying this IMT at heating in shock-wave or diamond anvil cell (DAC) experiments that have not been considered previously in its theoretical assessments.

\textit{Models and calculations.} The main subject of our study is the process of transition of molecular hydrogen heated under pressure to the plasma-like state. The complex nature of many-body electron-ion interactions requires models that are able to take into account possible non-adiabatic effects. For this purpose, in this work we consider the wave-packet molecular dynamics (WPMD) method using the electron force field (eFF) model~\cite{su2007} and the restricted open-shell Kohn-Sham (ROKS) DFT method for non-adiabatic \textit{ab initio} MD calculations with surface hopping (SH)~\cite{roks1998,cpmd2002}.

The WPMD Hamiltonian has terms that can be interpreted as kinetic energy contributions from ionic and electron degrees of freedom. Therefore, we are able to consider the temperatures of nuclear and electron subsystems independently. eFF provides a WPMD description of dense hydrogen~\cite{su2007,eff2019-Murillo} using a non-antisymmetric Hartree wavefunction for the many-electron system and representing nuclei as classical particles. A special term in the eFF Hamiltonian takes into account the energy contribution due to exchange interactions of electrons that makes eFF realistic but less computationally demanding approach than the fully anti-symmetrized WPMD formalism~\cite{Jakob-etal-WPMD_2009,Lavrinenko-etal-CPP2019}.

\begin{figure}
    \includegraphics[width=8.6cm]{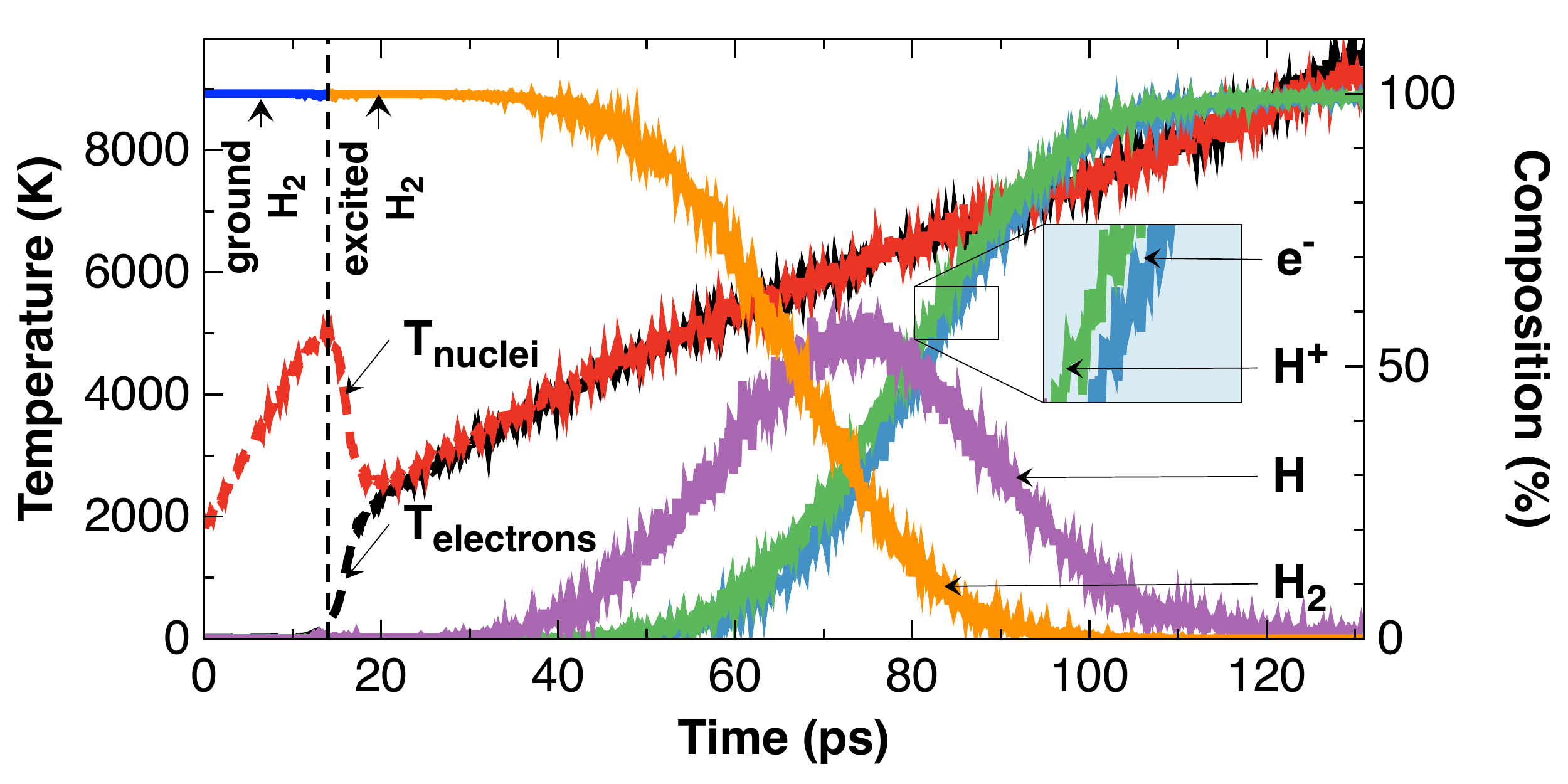}
    \caption{Temperatures and composition in the eFF model of fluid H$_2$ under isochoric heating ($\rho=0.3$~g/cc).
    \label{fig:isotopic_heating}}
\end{figure}

MD calculations with eFF are performed using LAMMPS~\cite{LAMMPS-1995} with periodic boundary conditions in a cubic simulation box. Isochoric heating is modelled by rescaling velocities of nuclei. Fig.~\ref{fig:isotopic_heating} shows the dependence of the nuclear and electronic temperatures on time during isochoric heating (0.09~K/fs) of 1000 H$_2$ molecules at $0.3$~g/cc (see the Supplemental Material~\cite{SupportingMaterial} for $0.6$~g/cc and other heating rates). The component analysis~\cite{HolstRedmerDesjarlais-PhysRevB.77.184201-2008} is performed along the MD trajectory. We see that initially the system evolves in its ground electronic state with the increasing temperature of molecules. Then, at each of the heating rates there are well defined ion temperatures when the electrons in molecules becomes excited. Shortly after, the ion temperature of the system gets to the new twice lower value that is the same both for electrons and for nuclei. The fact that the new ion temperature becomes 2x lower is determined by the artificial purely classical heat capacity of the eFF model system (since the number of degrees of freedom doubles). Problems with WP spreading~\cite{eff2019-Murillo} do not influence our results since such fast moving electrons appear only above $\sim 7000$~K.

Analysis of this excitation process shows that its nature corresponds to the non-adiabatic vibronic energy transfer from ionic to electronic degrees of freedom (see the animation in the Supplemental Material~\cite{SupportingMaterial}). eFF is able to give a qualitative description of excited H$_2$ molecules~\cite{SupportingMaterial} and we conclude that the formation of the excitonic molecular phase is observed. The component analysis shows that this phase remains stable under further isochoric heating. Fig.~\ref{fig:isotopic_heating} illustrates that in the eFF model dissociation proceeds gradually from this excitonic phase at higher temperatures. Here we should mention that an evident artifact of the eFF model is the irreversibility of the formation of this excitonic phase. Under cooling it does not become spontaneously the initial molecular phase in the ground state (with both electron WPs centred at the middle of each H$_2$ molecule).

The approximations assumed in eFF do not allow us to accept as necessarily realistic all the effects observed in the eFF MD simulations. But the formation of the excitonic molecular phase deserves careful consideration and trial by another less approximate non-adiabatic method.

\begin{figure}
    \includegraphics[width=8.0cm]{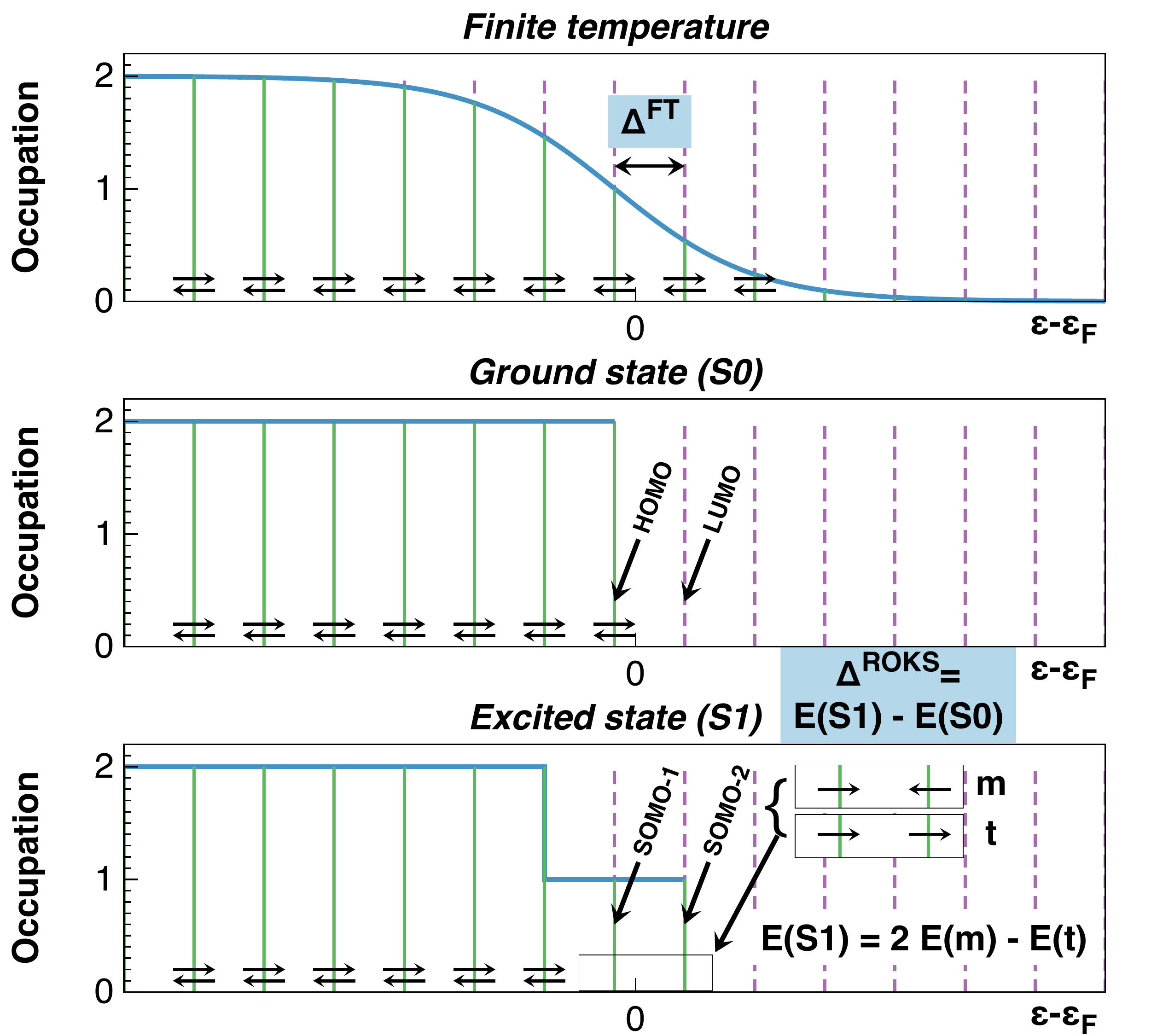}
    \caption{The scheme of the Kohn-Sham orbitals occupations in the DFT models for the finite temperature case, for the ground state case and for the ROKS case.\label{fig:occupation-schemes}}
\end{figure}

ROKS DFT allows us to calculate the electronic structure of the first singlet exited state (S1). We perform ROKS DFT calculations using CPMD~\cite{CPMD-v3.17} with the BLYP exchange-correlation functional, the Troullier-Martins pseudopotential for H with the 70 Ry plane wave cutoff and $\Gamma$-point sampling of the Brillouin zone.

Within the ROKS DFT model, it is assumed that the ground state of the many-electron system has been excited to the lowest excited state S1 via the transfer of a single electron from the HOMO orbital to the LUMO orbital. As a results, there are two singly occupied molecular orbitals in the system (SOMO-1/SOMO-2, see Fig.~\ref{fig:occupation-schemes}).

\begin{figure}
\includegraphics[width=8.2cm]{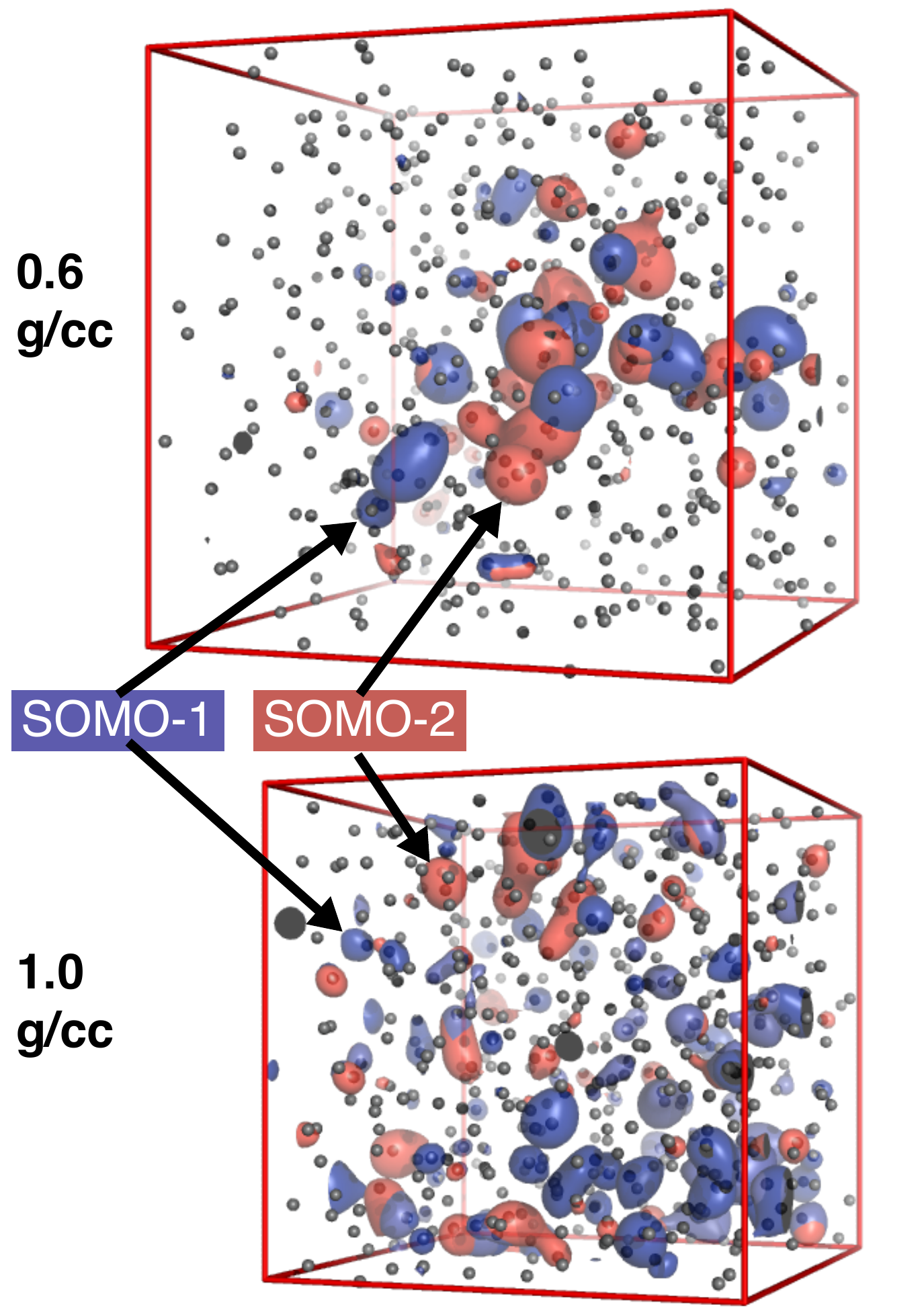}
\caption{Electron density distributions in the ROKS DFT model for the S0-equilibrated fluid structures at two densities (480 particles at 1300~K).\label{fig:representation_somo}}
\end{figure}

For each density and temperature considered the following equilibration procedure is implemented: initially a system with 60 atoms evolves in the electronic ground state with the Nose-Hoover thermostat for 100 thousand MD steps and then 200 thousand steps more are calculated in the NVE ensemble to ensure the stability of the temperature (the timestep equals 0.0234~fs). Then the system is replicated to 480 atoms, velocities are randomized and the system equilibrates for 7000 steps more. 

\begin{figure}
    \includegraphics[width=8.6cm]{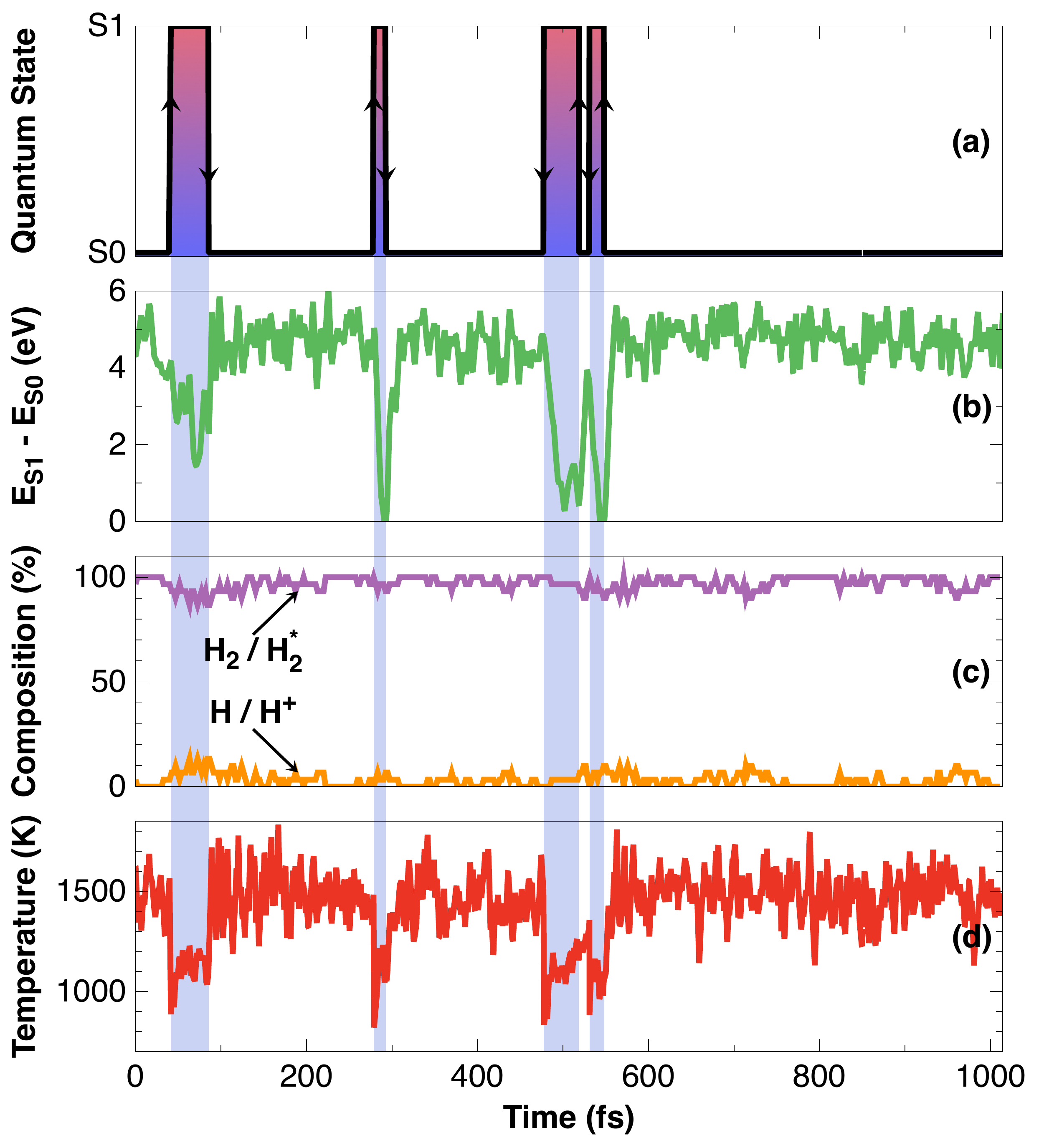}
    \caption{The parameters along the example ROKS DFT SH-MD trajectory for fluid H$_2$ (the state of the electronic subsystem, the S1-S0 energy gap, the concentrations of molecules and single atoms/ions, the temperature of ion subsystem).\label{fig:surface_hopping_dynamic}}
\end{figure}

It is instructive to analyze the shape of SOMOs in ROKS DFT for the equilibrated structures of fluid H$_2$. The shape of these orbitals shows the spatial extent of the excitons at different densities (Fig.~\ref{fig:representation_somo}). We see that at lower densities the S1 state is located on a cluster of molecules. The size of such a cluster becomes bigger at higher densities. In order to check the system size effects we have considered similar systems with 3840 atoms. Equilibration of these systems is performed in VASP~\cite{KresseHafner1993,KresseHafner1994,KresseFurthmuller1996,KresseFurthmuller1999}. The comparison of systems with 480 and 3840 atoms shows that the excitons are localized for densities up to 0.9~g/cc~\cite{SupportingMaterial}. The dependence of their visible spacial extension on the system size seems to be moderate. For higher densities the SOMO-1/2 orbitals cover the whole simulation box for both systems sizes that could be a sign of abrupt delocalization of electrons during IMT.

The excitonic phase in fluid hydrogen discussed here is not a completely new physical object. Excitons in liquids have been studied before, e.g. the results on the kinetics of formation of Xe$^*_2$ eximers have been reported~\cite{peterson1993dynamics}.

It is important to estimate typical lifetime values of this excitonic phase in fluid H$_2$. For this purpose, we use the surface hopping MD within the same ROKS DFT framework~\cite{cpmd2002}. From different initial conditions the ensemble of trajectories can be calculated starting from the ground state (S0) of the hydrogen system (0.6~g/cc, 1300~K). Along these trajectories the system can transfer spontaneously to S1 and after that it can return back to S0. These non-adiabatic transitions along MD trajectories are determined at each MD step by the transition probability calculated using the the wave-functions of S0 and S1 states for the current ionic configuration~\cite{cpmd2002}. Since the convergence of the system in S1 is much slower than in the ground state, we are able to use only the small system with 60 atoms for SH-MD calculations. The example of such a trajectory is shown on Fig.~\ref{fig:surface_hopping_dynamic}. Along this MD trajectory, we see 4 events of the spontaneous formation and recombination of excitons with lifetimes of 10-100~fs that is greater than the typical molecular vibration period (see~\cite{SupportingMaterial} for a lifetimes distribution).

\textit{Result (1): The vibronic mechanism of electronic transitions and exciton-like states in molecular fluid.} The calculations show the non-adibatic mechanism of energy transfer from molecular vibrations to electronic excitations. These excitations take place before dissociation of molecules into atoms. As a result, exciton-like states appear in molecular fluid H$_2$/D$_2$. These states are relatively stable with the lifetime values much larger than several molecular oscillations. The energy gap in these exciton-like states are smaller than in the ground state but still does not vanish. At the level of theory considered, these excitons are spatially localized at $\rho<0.9$~g/cc.

One important limitation of the ROKS DFT method is that it gives no possibility to access other excited states higher than S1. However even at this level of theory we see that the electron-ion dynamics during IMT in fluid hydrogen is characterized by the formation of relatively long-living excitonic structures. Here we should mention that even the proper description of the excited states of a single H$_2$ dimer is a complicated quantum mechanical problem that requires multi-reference methods~\cite{AstashkevichLavrov-ExcitedH2-lifetimes-2015}. ROKS DFT serves in this study as a computationally feasible approximation that is more realistic than the eFF model (eFF suffers from the lack of proper anti-symmetrization of the many-electron wavefunction and from the ambiguity of effective electron mass that affects the rates of non-adiabatic processes~\cite{Su-Goddard-AugerPNAS-2009}). This level of theory gives the possibility to justify at the semi-quantitative level the non-adiabatic nature of IMT and discuss the consequences of this thesis.

\textit{Result (2): The isotope effect.} The proposed non-adiabatic mechanism of IMT explains the strong isotopic effect observed in the DAC experiments for fluid H$_2$/D$_2$ ~\cite{ZaghooHusbandSilvera-PhysRevB.98.104102-2018}. Qualitatively, the probability of non-adiabatic transitions in a dimer is lower when the Massey parameter $\xi \sim \Delta / \dot{R}$ is higher ($\Delta$ is the energy gap and $\dot{R}$ is the relative velocity of atoms~\cite{doltsinis2002}). The H/D mass difference results in the significant difference of average atomic velocities in H$_2$/D$_2$. Therefore at the same temperature the probability of vibronic excitations in fluid D$_2$ is lower than in fluid H$_2$. The eFF calculations show the isotopic effect $\sim 400$~K~\cite{SupportingMaterial} that is close to experimental data~\cite{ZaghooHusbandSilvera-PhysRevB.98.104102-2018} and much larger than the isotopic effect based on NQE in equilibrium FPMD and QMC. The absence of the isotopic effect in the experimental results in DAC of Goncharov et al.~\cite{jiang2018insulatormetal} can be explained by measuring fluid H$_2$/D$_2$ properties during $\mu$s-long cooling contrary the DAC experiments of Silvera et al.~\cite{ZaghooHusbandSilvera-PhysRevB.98.104102-2018} that reports measurements done during heating. The relaxation of atomic ionized fluid at cooling to the state with molecules in their electronic ground state differs from vibronic molecular excitations at heating. At cooling, non-adiabatic radiationless internal conversion can not be expected to play a major role. One can expect that NQE provide the tunneling contribution increasing the non-adiabatic transitions rates. The methods that combine NQE with non-adiabatic dynamics are under development~\cite{Shushkov-etal-JCP-2012,markland2018nuclear}.

\textit{Result (3): The latent heat of transition.} During the ROKS DFT SH-MD runs the total energy of the system is conserved. It is implemented in CPMD via velocity rescaling of nuclei in the direction of the nonadiabatic coupling vector that compensates the energy changes in the electronic subsystem caused by S0-S1 and S1-S0 transitions. This means that after a vibronic excitation from S0 to S1 the system lowers its temperature (see Fig.~\ref{fig:surface_hopping_dynamic}). Such a cooling effect can help to stabilize the excitonic phase. Moreover, the $(E_{S1}-E_{S0})$ gap in the S1 state after initial relaxation can serve as an estimate for the detected latent heat of IMT reported in the shock-wave experiments~\cite{Holmes-etal-PhysRevB.52.15835-1995} and recently in the DAC experiments~\cite{houtput2019finite_PRB}: the analysis of experiments gives the value of 1-2~eV/atom and the example of Fig.~\ref{fig:surface_hopping_dynamic} corresponds to $(E_{S1}-E_{S0}) \sim 2$~eV that results in $\sim 1$~eV/atom and matches experimental values much better than the latent heat values deduced from the equilibrium DFT and QMC calculations of IMT ($\sim 0.04-0.05$~eV/atom~\cite{Morales-etal-PNAS-2010}). The proper comparison should be made for more specific density and temperature values. Besides, ROKS DFT is known to lower S0-S1 energy gaps~\cite{roks1998}. 

\textit{Discussion.} Experiments on IMT in fluid hydrogen are usually considered in the framework of the band gap closure mechanism. This explanation works well for semiconductors that are solids with nearly static ionic structure forming well-defined stable bands of electron states. Contrarily, fluid hydrogen is a state of matter with highly dynamic ionic structure. And we can speak of its electronic band structure only assuming an averaging over an ensemble of independent ionic configurations. This is the fundamental assumption that stands behind the FPMD/QMC based DFT calculations of conductivity, reflection and absorption coefficients~\cite{Rillo9770}.

The possible existence of the excitonic phase makes such an averaging questionable at the timescale of exciton lifetimes. We can hypothesise that the absorption of fluid H$_2$ during heating proceeds via formation of separated excitons. Later these excitons form clusters comparable with the probing pulse wavelength that results in reflectivity increase. Due to the finite time of exciton cluster nucleation and growth, the experimentally observed IMT temperatures could depend significantly on the corresponding heating rate. Therefore, there could be two reasons for the onset of reflectivity: the first is the formation of excitonic clusters and the second is the dissociation of molecules (both in the ground state phase and in the exciton clusters) and ionization.  Different timescales in different experiments could be the reasons of different data on the onset of absorption and reflectivity increase (e.g. $\sim 3$~ns for probing wide temperature range in the NIF experiments~\cite{Celliers2018} and $\sim 200$~ns for each temperature in the DAC experiments~\cite{ZaghooHusbandSilvera-PhysRevB.98.104102-2018}). This IMT picture can supplement the analysis of dynamic conductivity in dense fluid hydrogen~\cite{Zaghoo-PhysRevE.97.043205-2018} where the author concluded `that the non-free-electron nature of the
fluid could be explained by an increasing atomic polarizability': exciton clusters fit well to this description.

Recently, the intriguing results have been published that reveal metastable states in fluid H$_2$~\cite{NormanSaitovSartan-CPP-2019}. The nature of these metastable states is not clear since they were found using the equilibrium FT-DFT approach. However the inclination of fluid H$_2$ to the formation of exitonic phase near IMT can explain this effect as a result of a specific averaging in FT-DFT~\cite{StegailovZhilyaev-MolPhys-2016}. The excitonic phase suggested in this work should be a metastable phase. Therefore the transition between two molecular phases can be envisaged: between the fluid of molecules in the ground state and the growing fluid clusters of excited molecules. This fact can qualitatively explain results of~\cite{NormanSaitovSartan-CPP-2019}. Thus, the concept of the plasma phase transition~\cite{norman1968insufficiency,NormanSaitovStegailov-PPT-CPP-2015,NormanSaitov-CPP-PPT50-2019} should be probably reserved for the transition at ultrahigh temperature as has been suggested recently after careful equilibrium analysis of this IMT~\cite{Ackland-etal-PhysRevB.100.134109-2019}.  

The results presented point to a possibility of exciton formation in solid hydrogen~\cite{pickard2007structure,McMahonCeperly-PhysRevLett.106.165302-2011,DiasSilvera-Science-2017,Saitov-JETPLett-2019}, e.g. the black phase detected before metallization~\cite{DiasSilvera-Science-2017} could be attributed to formation of excitons that, however, are not able to form reflecting clusters. Excitons in crystals can be studied using well-established \textit{ab initio} techniques~\cite{rohlfing2000electron}. However, the coupling of electrons to lattice vibrations with NQE poses a significant theoretical challenge. 

\textit{Conclusions and Outlook.} The mechanism of the IMT in fluid H$_2$/D$_2$ has been studied beyond the Born-Oppenheimer approximation using non-abiabatic \textit{ab initio} methods. The possibility of formation of relatively long-living exciton-like states has been revealed. The proposed transition mechanism is the spontaneous vibronic excitation of molecules in fluid H$_2$/D$_2$ at heating. This mechanism gives an explanation of the isotopic difference of transition temperatures and of its detected latent heat. The proposed excitonic states in fluid H$_2$/D$_2$ can be similar to excitons in rare gas liquids~\cite{peterson1993dynamics}.

These non-adiabatic non-equilibrium effects are not expected to change the thermodynamic results predicted by equilibrium FPMD DFT and QMC calculations relevant to the interiors of giant planets. However as we have discussed, these effects are able to explain discrepancies of the results obtained in experimental studies of IMT in fluid H$_2$/D$_2$. Therefore, the important consequence is that, presumably, these experiments should not be interpreted using purely equilibrium theories.

\textit{Acknowledgments.} The authors thank Prof.~Genri Norman for drawing attention to this problem. V.S. is grateful to Prof.~Nikos Doltsinis for the introduction to the ROKS DFT SH-MD implementation in CPMD. The calculations were performed in the supercomputer centres of JIHT RAS, MIPT and NRU HSE. 


\begin{thebibliography}{77}%
\makeatletter
\providecommand \@ifxundefined [1]{%
 \@ifx{#1\undefined}
}%
\providecommand \@ifnum [1]{%
 \ifnum #1\expandafter \@firstoftwo
 \else \expandafter \@secondoftwo
 \fi
}%
\providecommand \@ifx [1]{%
 \ifx #1\expandafter \@firstoftwo
 \else \expandafter \@secondoftwo
 \fi
}%
\providecommand \natexlab [1]{#1}%
\providecommand \enquote  [1]{``#1''}%
\providecommand \bibnamefont  [1]{#1}%
\providecommand \bibfnamefont [1]{#1}%
\providecommand \citenamefont [1]{#1}%
\providecommand \href@noop [0]{\@secondoftwo}%
\providecommand \href [0]{\begingroup \@sanitize@url \@href}%
\providecommand \@href[1]{\@@startlink{#1}\@@href}%
\providecommand \@@href[1]{\endgroup#1\@@endlink}%
\providecommand \@sanitize@url [0]{\catcode `\\12\catcode `\$12\catcode
  `\&12\catcode `\#12\catcode `\^12\catcode `\_12\catcode `\%12\relax}%
\providecommand \@@startlink[1]{}%
\providecommand \@@endlink[0]{}%
\providecommand \url  [0]{\begingroup\@sanitize@url \@url }%
\providecommand \@url [1]{\endgroup\@href {#1}{\urlprefix }}%
\providecommand \urlprefix  [0]{URL }%
\providecommand \Eprint [0]{\href }%
\providecommand \doibase [0]{https://doi.org/}%
\providecommand \selectlanguage [0]{\@gobble}%
\providecommand \bibinfo  [0]{\@secondoftwo}%
\providecommand \bibfield  [0]{\@secondoftwo}%
\providecommand \translation [1]{[#1]}%
\providecommand \BibitemOpen [0]{}%
\providecommand \bibitemStop [0]{}%
\providecommand \bibitemNoStop [0]{.\EOS\space}%
\providecommand \EOS [0]{\spacefactor3000\relax}%
\providecommand \BibitemShut  [1]{\csname bibitem#1\endcsname}%
\let\auto@bib@innerbib\@empty
\bibitem [{\citenamefont {McMahon}\ \emph {et~al.}(2012)\citenamefont
  {McMahon}, \citenamefont {Morales}, \citenamefont {Pierleoni},\ and\
  \citenamefont {Ceperley}}]{mcmahon2012properties}%
  \BibitemOpen
  \bibfield  {author} {\bibinfo {author} {\bibfnamefont {J.~M.}\ \bibnamefont
  {McMahon}}, \bibinfo {author} {\bibfnamefont {M.~A.}\ \bibnamefont
  {Morales}}, \bibinfo {author} {\bibfnamefont {C.}~\bibnamefont {Pierleoni}},\
  and\ \bibinfo {author} {\bibfnamefont {D.~M.}\ \bibnamefont {Ceperley}},\
  }\bibfield  {title} {\bibinfo {title} {The properties of hydrogen and helium
  under extreme conditions},\ }\href
  {https://doi.org/10.1103/RevModPhys.84.1607} {\bibfield  {journal} {\bibinfo
  {journal} {Rev. Mod. Phys.}\ }\textbf {\bibinfo {volume} {84}},\ \bibinfo
  {pages} {1607} (\bibinfo {year} {2012})}\BibitemShut {NoStop}%
\bibitem [{\citenamefont {Utyuzh}\ and\ \citenamefont
  {Mikheyenkov}(2017)}]{Utyuzh_2017}%
  \BibitemOpen
  \bibfield  {author} {\bibinfo {author} {\bibfnamefont {A.~N.}\ \bibnamefont
  {Utyuzh}}\ and\ \bibinfo {author} {\bibfnamefont {A.~V.}\ \bibnamefont
  {Mikheyenkov}},\ }\bibfield  {title} {\bibinfo {title} {Hydrogen and its
  compounds under extreme pressure},\ }\href
  {https://doi.org/10.3367/ufne.2017.02.038077} {\bibfield  {journal} {\bibinfo
   {journal} {Physics-Uspekhi}\ }\textbf {\bibinfo {volume} {60}},\ \bibinfo
  {pages} {886} (\bibinfo {year} {2017})}\BibitemShut {NoStop}%
\bibitem [{\citenamefont {Weir}\ \emph {et~al.}(1996)\citenamefont {Weir},
  \citenamefont {Mitchell},\ and\ \citenamefont
  {Nellis}}]{Weir-etal-PhysRevLett.76.1860-1996}%
  \BibitemOpen
  \bibfield  {author} {\bibinfo {author} {\bibfnamefont {S.~T.}\ \bibnamefont
  {Weir}}, \bibinfo {author} {\bibfnamefont {A.~C.}\ \bibnamefont {Mitchell}},\
  and\ \bibinfo {author} {\bibfnamefont {W.~J.}\ \bibnamefont {Nellis}},\
  }\bibfield  {title} {\bibinfo {title} {Metallization of fluid molecular
  hydrogen at 140 gpa (1.4 mbar)},\ }\href
  {https://doi.org/10.1103/PhysRevLett.76.1860} {\bibfield  {journal} {\bibinfo
   {journal} {Phys. Rev. Lett.}\ }\textbf {\bibinfo {volume} {76}},\ \bibinfo
  {pages} {1860} (\bibinfo {year} {1996})}\BibitemShut {NoStop}%
\bibitem [{\citenamefont {Fortov}\ \emph {et~al.}(2007)\citenamefont {Fortov},
  \citenamefont {Ilkaev}, \citenamefont {Arinin}, \citenamefont {Burtzev},
  \citenamefont {Golubev}, \citenamefont {Iosilevskiy}, \citenamefont
  {Khrustalev}, \citenamefont {Mikhailov}, \citenamefont {Mochalov},
  \citenamefont {Ternovoi},\ and\ \citenamefont
  {Zhernokletov}}]{Fortov-etal-PhysRevLett.99.185001-2007}%
  \BibitemOpen
  \bibfield  {author} {\bibinfo {author} {\bibfnamefont {V.~E.}\ \bibnamefont
  {Fortov}}, \bibinfo {author} {\bibfnamefont {R.~I.}\ \bibnamefont {Ilkaev}},
  \bibinfo {author} {\bibfnamefont {V.~A.}\ \bibnamefont {Arinin}}, \bibinfo
  {author} {\bibfnamefont {V.~V.}\ \bibnamefont {Burtzev}}, \bibinfo {author}
  {\bibfnamefont {V.~A.}\ \bibnamefont {Golubev}}, \bibinfo {author}
  {\bibfnamefont {I.~L.}\ \bibnamefont {Iosilevskiy}}, \bibinfo {author}
  {\bibfnamefont {V.~V.}\ \bibnamefont {Khrustalev}}, \bibinfo {author}
  {\bibfnamefont {A.~L.}\ \bibnamefont {Mikhailov}}, \bibinfo {author}
  {\bibfnamefont {M.~A.}\ \bibnamefont {Mochalov}}, \bibinfo {author}
  {\bibfnamefont {V.~Y.}\ \bibnamefont {Ternovoi}},\ and\ \bibinfo {author}
  {\bibfnamefont {M.~V.}\ \bibnamefont {Zhernokletov}},\ }\bibfield  {title}
  {\bibinfo {title} {Phase transition in a strongly nonideal deuterium plasma
  generated by quasi-isentropical compression at megabar pressures},\ }\href
  {https://doi.org/10.1103/PhysRevLett.99.185001} {\bibfield  {journal}
  {\bibinfo  {journal} {Phys. Rev. Lett.}\ }\textbf {\bibinfo {volume} {99}},\
  \bibinfo {pages} {185001} (\bibinfo {year} {2007})}\BibitemShut {NoStop}%
\bibitem [{\citenamefont {Knudson}\ \emph {et~al.}(2015)\citenamefont
  {Knudson}, \citenamefont {Desjarlais}, \citenamefont {Becker}, \citenamefont
  {Lemke}, \citenamefont {Cochrane}, \citenamefont {Savage}, \citenamefont
  {Bliss}, \citenamefont {Mattsson},\ and\ \citenamefont
  {Redmer}}]{Knudson1455}%
  \BibitemOpen
  \bibfield  {author} {\bibinfo {author} {\bibfnamefont {M.~D.}\ \bibnamefont
  {Knudson}}, \bibinfo {author} {\bibfnamefont {M.~P.}\ \bibnamefont
  {Desjarlais}}, \bibinfo {author} {\bibfnamefont {A.}~\bibnamefont {Becker}},
  \bibinfo {author} {\bibfnamefont {R.~W.}\ \bibnamefont {Lemke}}, \bibinfo
  {author} {\bibfnamefont {K.~R.}\ \bibnamefont {Cochrane}}, \bibinfo {author}
  {\bibfnamefont {M.~E.}\ \bibnamefont {Savage}}, \bibinfo {author}
  {\bibfnamefont {D.~E.}\ \bibnamefont {Bliss}}, \bibinfo {author}
  {\bibfnamefont {T.~R.}\ \bibnamefont {Mattsson}},\ and\ \bibinfo {author}
  {\bibfnamefont {R.}~\bibnamefont {Redmer}},\ }\bibfield  {title} {\bibinfo
  {title} {Direct observation of an abrupt insulator-to-metal transition in
  dense liquid deuterium},\ }\href {https://doi.org/10.1126/science.aaa7471}
  {\bibfield  {journal} {\bibinfo  {journal} {Science}\ }\textbf {\bibinfo
  {volume} {348}},\ \bibinfo {pages} {1455} (\bibinfo {year}
  {2015})}\BibitemShut {NoStop}%
\bibitem [{\citenamefont {Mochalov}\ \emph {et~al.}(2017)\citenamefont
  {Mochalov}, \citenamefont {Il'kaev}, \citenamefont {Fortov}, \citenamefont
  {Mikhailov}, \citenamefont {Blikov}, \citenamefont {Ogorodnikov},
  \citenamefont {Gryaznov},\ and\ \citenamefont {Iosilevskii}}]{Mochalov2017}%
  \BibitemOpen
  \bibfield  {author} {\bibinfo {author} {\bibfnamefont {M.~A.}\ \bibnamefont
  {Mochalov}}, \bibinfo {author} {\bibfnamefont {R.~I.}\ \bibnamefont
  {Il'kaev}}, \bibinfo {author} {\bibfnamefont {V.~E.}\ \bibnamefont {Fortov}},
  \bibinfo {author} {\bibfnamefont {A.~L.}\ \bibnamefont {Mikhailov}}, \bibinfo
  {author} {\bibfnamefont {A.~O.}\ \bibnamefont {Blikov}}, \bibinfo {author}
  {\bibfnamefont {V.~A.}\ \bibnamefont {Ogorodnikov}}, \bibinfo {author}
  {\bibfnamefont {V.~K.}\ \bibnamefont {Gryaznov}},\ and\ \bibinfo {author}
  {\bibfnamefont {I.~L.}\ \bibnamefont {Iosilevskii}},\ }\bibfield  {title}
  {\bibinfo {title} {Quasi-isentropic compressibility of a strongly nonideal
  deuterium plasma at pressures of up to 5500 {GPa}: Nonideality and degeneracy
  effects},\ }\href {https://doi.org/10.1134/S1063776117020157} {\bibfield
  {journal} {\bibinfo  {journal} {Journal of Experimental and Theoretical
  Physics}\ }\textbf {\bibinfo {volume} {124}},\ \bibinfo {pages} {505}
  (\bibinfo {year} {2017})}\BibitemShut {NoStop}%
\bibitem [{\citenamefont {Celliers}\ \emph {et~al.}(2018)\citenamefont
  {Celliers}, \citenamefont {Millot}, \citenamefont {Brygoo}, \citenamefont
  {McWilliams}, \citenamefont {Fratanduono}, \citenamefont {Rygg},
  \citenamefont {Goncharov}, \citenamefont {Loubeyre}, \citenamefont {Eggert},
  \citenamefont {Peterson}, \citenamefont {Meezan}, \citenamefont {Le~Pape},
  \citenamefont {Collins}, \citenamefont {Jeanloz},\ and\ \citenamefont
  {Hemley}}]{Celliers2018}%
  \BibitemOpen
  \bibfield  {author} {\bibinfo {author} {\bibfnamefont {P.~M.}\ \bibnamefont
  {Celliers}}, \bibinfo {author} {\bibfnamefont {M.}~\bibnamefont {Millot}},
  \bibinfo {author} {\bibfnamefont {S.}~\bibnamefont {Brygoo}}, \bibinfo
  {author} {\bibfnamefont {R.~S.}\ \bibnamefont {McWilliams}}, \bibinfo
  {author} {\bibfnamefont {D.~E.}\ \bibnamefont {Fratanduono}}, \bibinfo
  {author} {\bibfnamefont {J.~R.}\ \bibnamefont {Rygg}}, \bibinfo {author}
  {\bibfnamefont {A.~F.}\ \bibnamefont {Goncharov}}, \bibinfo {author}
  {\bibfnamefont {P.}~\bibnamefont {Loubeyre}}, \bibinfo {author}
  {\bibfnamefont {J.~H.}\ \bibnamefont {Eggert}}, \bibinfo {author}
  {\bibfnamefont {J.~L.}\ \bibnamefont {Peterson}}, \bibinfo {author}
  {\bibfnamefont {N.~B.}\ \bibnamefont {Meezan}}, \bibinfo {author}
  {\bibfnamefont {S.}~\bibnamefont {Le~Pape}}, \bibinfo {author} {\bibfnamefont
  {G.~W.}\ \bibnamefont {Collins}}, \bibinfo {author} {\bibfnamefont
  {R.}~\bibnamefont {Jeanloz}},\ and\ \bibinfo {author} {\bibfnamefont {R.~J.}\
  \bibnamefont {Hemley}},\ }\bibfield  {title} {\bibinfo {title}
  {Insulator-metal transition in dense fluid deuterium},\ }\href
  {https://doi.org/10.1126/science.aat0970} {\bibfield  {journal} {\bibinfo
  {journal} {Science}\ }\textbf {\bibinfo {volume} {361}},\ \bibinfo {pages}
  {677} (\bibinfo {year} {2018})}\BibitemShut {NoStop}%
\bibitem [{\citenamefont {Loubeyre}\ \emph {et~al.}(2004)\citenamefont
  {Loubeyre}, \citenamefont {Celliers}, \citenamefont {Hicks}, \citenamefont
  {Henry}, \citenamefont {Dewaele}, \citenamefont {Pasley}, \citenamefont
  {Eggert}, \citenamefont {Koenig}, \citenamefont {Occelli}, \citenamefont
  {Lee}, \citenamefont {Jeanloz}, \citenamefont {Neely}, \citenamefont
  {Benuzzi-Mounaix}, \citenamefont {Bradley}, \citenamefont {Bastea},
  \citenamefont {Moon},\ and\ \citenamefont
  {Collins}}]{Loubeyre-etal-HighPressRes-2004}%
  \BibitemOpen
  \bibfield  {author} {\bibinfo {author} {\bibfnamefont {P.}~\bibnamefont
  {Loubeyre}}, \bibinfo {author} {\bibfnamefont {P.~M.}\ \bibnamefont
  {Celliers}}, \bibinfo {author} {\bibfnamefont {D.~G.}\ \bibnamefont {Hicks}},
  \bibinfo {author} {\bibfnamefont {E.}~\bibnamefont {Henry}}, \bibinfo
  {author} {\bibfnamefont {A.}~\bibnamefont {Dewaele}}, \bibinfo {author}
  {\bibfnamefont {J.}~\bibnamefont {Pasley}}, \bibinfo {author} {\bibfnamefont
  {J.}~\bibnamefont {Eggert}}, \bibinfo {author} {\bibfnamefont
  {M.}~\bibnamefont {Koenig}}, \bibinfo {author} {\bibfnamefont
  {F.}~\bibnamefont {Occelli}}, \bibinfo {author} {\bibfnamefont {K.~M.}\
  \bibnamefont {Lee}}, \bibinfo {author} {\bibfnamefont {R.}~\bibnamefont
  {Jeanloz}}, \bibinfo {author} {\bibfnamefont {D.}~\bibnamefont {Neely}},
  \bibinfo {author} {\bibfnamefont {A.}~\bibnamefont {Benuzzi-Mounaix}},
  \bibinfo {author} {\bibfnamefont {D.}~\bibnamefont {Bradley}}, \bibinfo
  {author} {\bibfnamefont {M.}~\bibnamefont {Bastea}}, \bibinfo {author}
  {\bibfnamefont {S.}~\bibnamefont {Moon}},\ and\ \bibinfo {author}
  {\bibfnamefont {G.~W.}\ \bibnamefont {Collins}},\ }\bibfield  {title}
  {\bibinfo {title} {Coupling static and dynamic compressions: first
  measurements in dense hydrogen},\ }\href
  {https://doi.org/10.1080/08957950310001635792} {\bibfield  {journal}
  {\bibinfo  {journal} {High Pressure Research}\ }\textbf {\bibinfo {volume}
  {24}},\ \bibinfo {pages} {25} (\bibinfo {year} {2004})}\BibitemShut {NoStop}%
\bibitem [{\citenamefont {Dzyabura}\ \emph {et~al.}(2013)\citenamefont
  {Dzyabura}, \citenamefont {Zaghoo},\ and\ \citenamefont
  {Silvera}}]{Dzyabura2013}%
  \BibitemOpen
  \bibfield  {author} {\bibinfo {author} {\bibfnamefont {V.}~\bibnamefont
  {Dzyabura}}, \bibinfo {author} {\bibfnamefont {M.}~\bibnamefont {Zaghoo}},\
  and\ \bibinfo {author} {\bibfnamefont {I.~F.}\ \bibnamefont {Silvera}},\
  }\bibfield  {title} {\bibinfo {title} {Evidence of a
  liquid{\textendash}liquid phase transition in hot dense hydrogen},\ }\href
  {https://doi.org/10.1073/pnas.1300718110} {\bibfield  {journal} {\bibinfo
  {journal} {Proceedings of the National Academy of Sciences}\ }\textbf
  {\bibinfo {volume} {110}},\ \bibinfo {pages} {8040} (\bibinfo {year}
  {2013})}\BibitemShut {NoStop}%
\bibitem [{\citenamefont {Ohta}\ \emph {et~al.}(2015)\citenamefont {Ohta},
  \citenamefont {Ichimaru}, \citenamefont {Einaga}, \citenamefont {Kawaguchi},
  \citenamefont {Shimizu}, \citenamefont {Matsuoka}, \citenamefont {Hirao},\
  and\ \citenamefont {Ohishi}}]{ohta2015phase}%
  \BibitemOpen
  \bibfield  {author} {\bibinfo {author} {\bibfnamefont {K.}~\bibnamefont
  {Ohta}}, \bibinfo {author} {\bibfnamefont {K.}~\bibnamefont {Ichimaru}},
  \bibinfo {author} {\bibfnamefont {M.}~\bibnamefont {Einaga}}, \bibinfo
  {author} {\bibfnamefont {S.}~\bibnamefont {Kawaguchi}}, \bibinfo {author}
  {\bibfnamefont {K.}~\bibnamefont {Shimizu}}, \bibinfo {author} {\bibfnamefont
  {T.}~\bibnamefont {Matsuoka}}, \bibinfo {author} {\bibfnamefont
  {N.}~\bibnamefont {Hirao}},\ and\ \bibinfo {author} {\bibfnamefont
  {Y.}~\bibnamefont {Ohishi}},\ }\bibfield  {title} {\bibinfo {title} {Phase
  boundary of hot dense fluid hydrogen},\ }\href@noop {} {\bibfield  {journal}
  {\bibinfo  {journal} {Scientific reports}\ }\textbf {\bibinfo {volume} {5}},\
  \bibinfo {pages} {16560} (\bibinfo {year} {2015})}\BibitemShut {NoStop}%
\bibitem [{\citenamefont {McWilliams}\ \emph {et~al.}(2016)\citenamefont
  {McWilliams}, \citenamefont {Dalton}, \citenamefont {Mahmood},\ and\
  \citenamefont {Goncharov}}]{goncharov2016}%
  \BibitemOpen
  \bibfield  {author} {\bibinfo {author} {\bibfnamefont {R.~S.}\ \bibnamefont
  {McWilliams}}, \bibinfo {author} {\bibfnamefont {D.~A.}\ \bibnamefont
  {Dalton}}, \bibinfo {author} {\bibfnamefont {M.~F.}\ \bibnamefont
  {Mahmood}},\ and\ \bibinfo {author} {\bibfnamefont {A.~F.}\ \bibnamefont
  {Goncharov}},\ }\bibfield  {title} {\bibinfo {title} {Optical properties of
  fluid hydrogen at the transition to a conducting state},\ }\href
  {https://doi.org/10.1103/PhysRevLett.116.255501} {\bibfield  {journal}
  {\bibinfo  {journal} {Phys. Rev. Lett.}\ }\textbf {\bibinfo {volume} {116}},\
  \bibinfo {pages} {255501} (\bibinfo {year} {2016})}\BibitemShut {NoStop}%
\bibitem [{\citenamefont {Zaghoo}\ \emph {et~al.}(2018)\citenamefont {Zaghoo},
  \citenamefont {Husband},\ and\ \citenamefont
  {Silvera}}]{ZaghooHusbandSilvera-PhysRevB.98.104102-2018}%
  \BibitemOpen
  \bibfield  {author} {\bibinfo {author} {\bibfnamefont {M.}~\bibnamefont
  {Zaghoo}}, \bibinfo {author} {\bibfnamefont {R.~J.}\ \bibnamefont
  {Husband}},\ and\ \bibinfo {author} {\bibfnamefont {I.~F.}\ \bibnamefont
  {Silvera}},\ }\bibfield  {title} {\bibinfo {title} {Striking isotope effect
  on the metallization phase lines of liquid hydrogen and deuterium},\ }\href
  {https://doi.org/10.1103/PhysRevB.98.104102} {\bibfield  {journal} {\bibinfo
  {journal} {Phys. Rev. B}\ }\textbf {\bibinfo {volume} {98}},\ \bibinfo
  {pages} {104102} (\bibinfo {year} {2018})}\BibitemShut {NoStop}%
\bibitem [{\citenamefont {Norman}\ and\ \citenamefont
  {Starostin}(1968)}]{norman1968insufficiency}%
  \BibitemOpen
  \bibfield  {author} {\bibinfo {author} {\bibfnamefont {G.}~\bibnamefont
  {Norman}}\ and\ \bibinfo {author} {\bibfnamefont {A.}~\bibnamefont
  {Starostin}},\ }\bibfield  {title} {\bibinfo {title} {Insufficiency of the
  classical description of a nondegenerate dense plasma},\ }\href@noop {}
  {\bibfield  {journal} {\bibinfo  {journal} {High Temperature}\ }\textbf
  {\bibinfo {volume} {6}},\ \bibinfo {pages} {394} (\bibinfo {year}
  {1968})}\BibitemShut {NoStop}%
\bibitem [{\citenamefont {Biberman}\ and\ \citenamefont
  {Norman}(1969)}]{Biberman1969}%
  \BibitemOpen
  \bibfield  {author} {\bibinfo {author} {\bibfnamefont {L.~M.}\ \bibnamefont
  {Biberman}}\ and\ \bibinfo {author} {\bibfnamefont {G.~E.}\ \bibnamefont
  {Norman}},\ }\bibfield  {title} {\bibinfo {title} {On existence possibility
  of overcooled dense plasma},\ }\href@noop {} {\bibfield  {journal} {\bibinfo
  {journal} {High Temp.}\ }\textbf {\bibinfo {volume} {7}},\ \bibinfo {pages}
  {822} (\bibinfo {year} {1969})}\BibitemShut {NoStop}%
\bibitem [{\citenamefont {Lebowitz}\ and\ \citenamefont
  {Lieb}(1969)}]{Lebowitz1969}%
  \BibitemOpen
  \bibfield  {author} {\bibinfo {author} {\bibfnamefont {J.~L.}\ \bibnamefont
  {Lebowitz}}\ and\ \bibinfo {author} {\bibfnamefont {E.~H.}\ \bibnamefont
  {Lieb}},\ }\bibfield  {title} {\bibinfo {title} {Existence of thermodynamics
  for real matter with coulomb forces},\ }\href
  {https://doi.org/10.1103/PhysRevLett.22.631} {\bibfield  {journal} {\bibinfo
  {journal} {Phys. Rev. Lett.}\ }\textbf {\bibinfo {volume} {22}},\ \bibinfo
  {pages} {631} (\bibinfo {year} {1969})}\BibitemShut {NoStop}%
\bibitem [{\citenamefont {{Norman}}\ and\ \citenamefont
  {{Starostin}}(1970)}]{Norman1970}%
  \BibitemOpen
  \bibfield  {author} {\bibinfo {author} {\bibfnamefont {G.~{\'E}.}\
  \bibnamefont {{Norman}}}\ and\ \bibinfo {author} {\bibfnamefont {A.~N.}\
  \bibnamefont {{Starostin}}},\ }\bibfield  {title} {\bibinfo {title}
  {{Thermodynamics of a dense plasma}},\ }\href
  {https://doi.org/10.1007/BF00607515} {\bibfield  {journal} {\bibinfo
  {journal} {J. Appl. Spectrosc.}\ }\textbf {\bibinfo {volume} {13}},\ \bibinfo
  {pages} {965} (\bibinfo {year} {1970})}\BibitemShut {NoStop}%
\bibitem [{\citenamefont {Norman}\ and\ \citenamefont
  {Starostin}(1970)}]{Norman1970a}%
  \BibitemOpen
  \bibfield  {author} {\bibinfo {author} {\bibfnamefont {G.~E.}\ \bibnamefont
  {Norman}}\ and\ \bibinfo {author} {\bibfnamefont {A.~N.}\ \bibnamefont
  {Starostin}},\ }\bibfield  {title} {\bibinfo {title} {Thermodynamics of a
  strongly nonideal plasma},\ }\href@noop {} {\bibfield  {journal} {\bibinfo
  {journal} {High Temp.}\ }\textbf {\bibinfo {volume} {8}},\ \bibinfo {pages}
  {381} (\bibinfo {year} {1970})}\BibitemShut {NoStop}%
\bibitem [{\citenamefont {Ebeling}(1971)}]{Ebeling1971}%
  \BibitemOpen
  \bibfield  {author} {\bibinfo {author} {\bibfnamefont {W.}~\bibnamefont
  {Ebeling}},\ }\bibfield  {title} {\bibinfo {title} {Quantum statistics of
  ionization and shielding effects in non-degenerate moderately doped
  semiconductors},\ }\href {https://doi.org/10.1002/pssb.2220460122} {\bibfield
   {journal} {\bibinfo  {journal} {Phys. Stat. Sol. B}\ }\textbf {\bibinfo
  {volume} {46}},\ \bibinfo {pages} {243} (\bibinfo {year} {1971})}\BibitemShut
  {NoStop}%
\bibitem [{\citenamefont {Kraeft}\ \emph {et~al.}(1986)\citenamefont {Kraeft},
  \citenamefont {Kremp}, \citenamefont {Ebeling},\ and\ \citenamefont
  {R{\"o}pke}}]{Kraeft1986}%
  \BibitemOpen
  \bibfield  {author} {\bibinfo {author} {\bibfnamefont {W.}~\bibnamefont
  {Kraeft}}, \bibinfo {author} {\bibfnamefont {D.}~\bibnamefont {Kremp}},
  \bibinfo {author} {\bibfnamefont {W.}~\bibnamefont {Ebeling}},\ and\ \bibinfo
  {author} {\bibfnamefont {G.}~\bibnamefont {R{\"o}pke}},\ }\href@noop {}
  {\emph {\bibinfo {title} {Quantum statistics of charged particle systems}}}\
  (\bibinfo  {publisher} {Springer},\ \bibinfo {year} {1986})\BibitemShut
  {NoStop}%
\bibitem [{\citenamefont {{Saumon}}\ and\ \citenamefont
  {{Chabrier}}(1991)}]{Saumon1991}%
  \BibitemOpen
  \bibfield  {author} {\bibinfo {author} {\bibfnamefont {D.}~\bibnamefont
  {{Saumon}}}\ and\ \bibinfo {author} {\bibfnamefont {G.}~\bibnamefont
  {{Chabrier}}},\ }\bibfield  {title} {\bibinfo {title} {{Fluid hydrogen at
  high density - Pressure dissociation}},\ }\href@noop {} {\bibfield  {journal}
  {\bibinfo  {journal} {Phys. Rev. A}\ }\textbf {\bibinfo {volume} {44}},\
  \bibinfo {pages} {5122} (\bibinfo {year} {1991})}\BibitemShut {NoStop}%
\bibitem [{\citenamefont {{Reinholz}}\ \emph {et~al.}(1995)\citenamefont
  {{Reinholz}}, \citenamefont {{Redmer}},\ and\ \citenamefont
  {{Nagel}}}]{Reinholz1995}%
  \BibitemOpen
  \bibfield  {author} {\bibinfo {author} {\bibfnamefont {H.}~\bibnamefont
  {{Reinholz}}}, \bibinfo {author} {\bibfnamefont {R.}~\bibnamefont
  {{Redmer}}},\ and\ \bibinfo {author} {\bibfnamefont {S.}~\bibnamefont
  {{Nagel}}},\ }\bibfield  {title} {\bibinfo {title} {{Thermodynamic and
  transport properties of dense hydrogen plasmas}},\ }\href@noop {} {\bibfield
  {journal} {\bibinfo  {journal} {Phys. Rev. E}\ }\textbf {\bibinfo {volume}
  {52}},\ \bibinfo {pages} {5368} (\bibinfo {year} {1995})}\BibitemShut
  {NoStop}%
\bibitem [{\citenamefont {Ebeling}\ and\ \citenamefont
  {Norman}(2003)}]{Ebeling2003}%
  \BibitemOpen
  \bibfield  {author} {\bibinfo {author} {\bibfnamefont {W.}~\bibnamefont
  {Ebeling}}\ and\ \bibinfo {author} {\bibfnamefont {G.}~\bibnamefont
  {Norman}},\ }\bibfield  {title} {\bibinfo {title} {Coulombic phase
  transitions in dense plasmas},\ }\href@noop {} {\bibfield  {journal}
  {\bibinfo  {journal} {J. Stat. Phys.}\ }\textbf {\bibinfo {volume} {110}},\
  \bibinfo {pages} {861} (\bibinfo {year} {2003})}\BibitemShut {NoStop}%
\bibitem [{\citenamefont {Norman}(2006)}]{Norman2006}%
  \BibitemOpen
  \bibfield  {author} {\bibinfo {author} {\bibfnamefont {G.~E.}\ \bibnamefont
  {Norman}},\ }\bibfield  {title} {\bibinfo {title} {Phase diagram of ultracold
  strongly coupled plasmas},\ }\href@noop {} {\bibfield  {journal} {\bibinfo
  {journal} {J. Phys. A: Math. Gen.}\ }\textbf {\bibinfo {volume} {39}},\
  \bibinfo {pages} {4579} (\bibinfo {year} {2006})}\BibitemShut {NoStop}%
\bibitem [{\citenamefont {{Khomkin}}\ and\ \citenamefont
  {{Shumikhin}}(2013)}]{Khomkin2013}%
  \BibitemOpen
  \bibfield  {author} {\bibinfo {author} {\bibfnamefont {A.~L.}\ \bibnamefont
  {{Khomkin}}}\ and\ \bibinfo {author} {\bibfnamefont {A.~S.}\ \bibnamefont
  {{Shumikhin}}},\ }\bibfield  {title} {\bibinfo {title} {{Phase transition
  into the metallic state in hypothetical (without molecules) dense atomic
  hydrogen}},\ }\href {https://doi.org/10.1134/S1063780X13100061} {\bibfield
  {journal} {\bibinfo  {journal} {Plasma Phys. Rep.}\ }\textbf {\bibinfo
  {volume} {39}},\ \bibinfo {pages} {857} (\bibinfo {year} {2013})}\BibitemShut
  {NoStop}%
\bibitem [{\citenamefont {Starostin}\ \emph {et~al.}(2016)\citenamefont
  {Starostin}, \citenamefont {Gryaznov},\ and\ \citenamefont
  {Filippov}}]{Starostin2016}%
  \BibitemOpen
  \bibfield  {author} {\bibinfo {author} {\bibfnamefont {A.~N.}\ \bibnamefont
  {Starostin}}, \bibinfo {author} {\bibfnamefont {V.~K.}\ \bibnamefont
  {Gryaznov}},\ and\ \bibinfo {author} {\bibfnamefont {A.~V.}\ \bibnamefont
  {Filippov}},\ }\bibfield  {title} {\bibinfo {title} {Thermoelectric
  properties of a plasma at megabar pressures},\ }\href@noop {} {\bibfield
  {journal} {\bibinfo  {journal} {J. Exp. Theor. Phys. Lett.}\ }\textbf
  {\bibinfo {volume} {104}},\ \bibinfo {pages} {696} (\bibinfo {year}
  {2016})}\BibitemShut {NoStop}%
\bibitem [{\citenamefont {Ebeling}\ \emph {et~al.}(2017)\citenamefont
  {Ebeling}, \citenamefont {Fortov},\ and\ \citenamefont
  {Filinov}}]{Ebeling2017}%
  \BibitemOpen
  \bibfield  {author} {\bibinfo {author} {\bibfnamefont {W.}~\bibnamefont
  {Ebeling}}, \bibinfo {author} {\bibfnamefont {V.}~\bibnamefont {Fortov}},\
  and\ \bibinfo {author} {\bibfnamefont {V.}~\bibnamefont {Filinov}},\
  }\href@noop {} {\emph {\bibinfo {title} {Quantum Statistics of Dense Gases
  and Nonideal Plasmas}}}\ (\bibinfo  {publisher} {Springer},\ \bibinfo {year}
  {2017})\BibitemShut {NoStop}%
\bibitem [{\citenamefont {Filinov}\ and\ \citenamefont
  {Norman}(1975)}]{FilinovNorman-PRA-1975}%
  \BibitemOpen
  \bibfield  {author} {\bibinfo {author} {\bibfnamefont {V.}~\bibnamefont
  {Filinov}}\ and\ \bibinfo {author} {\bibfnamefont {G.}~\bibnamefont
  {Norman}},\ }\bibfield  {title} {\bibinfo {title} {On phase transition in a
  non-ideal plasma},\ }\href@noop {} {\bibfield  {journal} {\bibinfo  {journal}
  {Physics Letters A}\ }\textbf {\bibinfo {volume} {55}},\ \bibinfo {pages}
  {219 } (\bibinfo {year} {1975})}\BibitemShut {NoStop}%
\bibitem [{\citenamefont {Redmer}\ and\ \citenamefont
  {Holst}(2010)}]{Redmer2010book}%
  \BibitemOpen
  \bibfield  {author} {\bibinfo {author} {\bibfnamefont {R.}~\bibnamefont
  {Redmer}}\ and\ \bibinfo {author} {\bibfnamefont {B.}~\bibnamefont {Holst}},\
  }\bibinfo {title} {Metal--insulator transition in dense hydrogen},\ in\ \href
  {https://doi.org/10.1007/978-3-642-03953-9_4} {\emph {\bibinfo {booktitle}
  {Metal-to-Nonmetal Transitions}}},\ \bibinfo {editor} {edited by\ \bibinfo
  {editor} {\bibfnamefont {R.}~\bibnamefont {Redmer}}, \bibinfo {editor}
  {\bibfnamefont {F.}~\bibnamefont {Hensel}},\ and\ \bibinfo {editor}
  {\bibfnamefont {B.}~\bibnamefont {Holst}}}\ (\bibinfo  {publisher} {Springer
  Berlin Heidelberg},\ \bibinfo {address} {Berlin, Heidelberg},\ \bibinfo
  {year} {2010})\ pp.\ \bibinfo {pages} {63--84}\BibitemShut {NoStop}%
\bibitem [{\citenamefont {Scandolo}(2003)}]{Scandolo-PNAS2003}%
  \BibitemOpen
  \bibfield  {author} {\bibinfo {author} {\bibfnamefont {S.}~\bibnamefont
  {Scandolo}},\ }\bibfield  {title} {\bibinfo {title}
  {Liquid{\textendash}liquid phase transition in compressed hydrogen from
  first-principles simulations},\ }\href@noop {} {\bibfield  {journal}
  {\bibinfo  {journal} {Proceedings of the National Academy of Sciences}\
  }\textbf {\bibinfo {volume} {100}},\ \bibinfo {pages} {3051} (\bibinfo {year}
  {2003})}\BibitemShut {NoStop}%
\bibitem [{\citenamefont {Tamblyn}\ and\ \citenamefont
  {Bonev}(2010)}]{TamblynBonev-PhysRevLett.104.065702-2010}%
  \BibitemOpen
  \bibfield  {author} {\bibinfo {author} {\bibfnamefont {I.}~\bibnamefont
  {Tamblyn}}\ and\ \bibinfo {author} {\bibfnamefont {S.~A.}\ \bibnamefont
  {Bonev}},\ }\bibfield  {title} {\bibinfo {title} {Structure and phase
  boundaries of compressed liquid hydrogen},\ }\href
  {https://doi.org/10.1103/PhysRevLett.104.065702} {\bibfield  {journal}
  {\bibinfo  {journal} {Phys. Rev. Lett.}\ }\textbf {\bibinfo {volume} {104}},\
  \bibinfo {pages} {065702} (\bibinfo {year} {2010})}\BibitemShut {NoStop}%
\bibitem [{\citenamefont {Morales}\ \emph {et~al.}(2013)\citenamefont
  {Morales}, \citenamefont {McMahon}, \citenamefont {Pierleoni},\ and\
  \citenamefont {Ceperley}}]{Morales-et-al-PhysRevLett.110.065702-2013}%
  \BibitemOpen
  \bibfield  {author} {\bibinfo {author} {\bibfnamefont {M.~A.}\ \bibnamefont
  {Morales}}, \bibinfo {author} {\bibfnamefont {J.~M.}\ \bibnamefont
  {McMahon}}, \bibinfo {author} {\bibfnamefont {C.}~\bibnamefont {Pierleoni}},\
  and\ \bibinfo {author} {\bibfnamefont {D.~M.}\ \bibnamefont {Ceperley}},\
  }\bibfield  {title} {\bibinfo {title} {Nuclear quantum effects and nonlocal
  exchange-correlation functionals applied to liquid hydrogen at high
  pressure},\ }\href {https://doi.org/10.1103/PhysRevLett.110.065702}
  {\bibfield  {journal} {\bibinfo  {journal} {Phys. Rev. Lett.}\ }\textbf
  {\bibinfo {volume} {110}},\ \bibinfo {pages} {065702} (\bibinfo {year}
  {2013})}\BibitemShut {NoStop}%
\bibitem [{\citenamefont {Mazzola}\ and\ \citenamefont
  {Sorella}(2017)}]{Mazolla2017}%
  \BibitemOpen
  \bibfield  {author} {\bibinfo {author} {\bibfnamefont {G.}~\bibnamefont
  {Mazzola}}\ and\ \bibinfo {author} {\bibfnamefont {S.}~\bibnamefont
  {Sorella}},\ }\bibfield  {title} {\bibinfo {title} {Accelerating ab initio
  molecular dynamics and probing the weak dispersive forces in dense liquid
  hydrogen},\ }\href {https://doi.org/10.1103/PhysRevLett.118.015703}
  {\bibfield  {journal} {\bibinfo  {journal} {Phys. Rev. Lett.}\ }\textbf
  {\bibinfo {volume} {118}},\ \bibinfo {pages} {015703} (\bibinfo {year}
  {2017})}\BibitemShut {NoStop}%
\bibitem [{\citenamefont {Knudson}\ and\ \citenamefont
  {Desjarlais}(2017)}]{Knudson-Desjarlais-PhysRevLett.118.035501-2017}%
  \BibitemOpen
  \bibfield  {author} {\bibinfo {author} {\bibfnamefont {M.~D.}\ \bibnamefont
  {Knudson}}\ and\ \bibinfo {author} {\bibfnamefont {M.~P.}\ \bibnamefont
  {Desjarlais}},\ }\bibfield  {title} {\bibinfo {title} {High-precision shock
  wave measurements of deuterium: Evaluation of exchange-correlation
  functionals at the molecular-to-atomic transition},\ }\href
  {https://doi.org/10.1103/PhysRevLett.118.035501} {\bibfield  {journal}
  {\bibinfo  {journal} {Phys. Rev. Lett.}\ }\textbf {\bibinfo {volume} {118}},\
  \bibinfo {pages} {035501} (\bibinfo {year} {2017})}\BibitemShut {NoStop}%
\bibitem [{\citenamefont {Knudson}\ \emph {et~al.}(2018)\citenamefont
  {Knudson}, \citenamefont {Desjarlais}, \citenamefont {Preising},\ and\
  \citenamefont {Redmer}}]{Knudson-etal-PhysRevB.98.174110-2018}%
  \BibitemOpen
  \bibfield  {author} {\bibinfo {author} {\bibfnamefont {M.~D.}\ \bibnamefont
  {Knudson}}, \bibinfo {author} {\bibfnamefont {M.~P.}\ \bibnamefont
  {Desjarlais}}, \bibinfo {author} {\bibfnamefont {M.}~\bibnamefont
  {Preising}},\ and\ \bibinfo {author} {\bibfnamefont {R.}~\bibnamefont
  {Redmer}},\ }\bibfield  {title} {\bibinfo {title} {Evaluation of
  exchange-correlation functionals with multiple-shock conductivity
  measurements in hydrogen and deuterium at the molecular-to-atomic
  transition},\ }\href {https://doi.org/10.1103/PhysRevB.98.174110} {\bibfield
  {journal} {\bibinfo  {journal} {Phys. Rev. B}\ }\textbf {\bibinfo {volume}
  {98}},\ \bibinfo {pages} {174110} (\bibinfo {year} {2018})}\BibitemShut
  {NoStop}%
\bibitem [{\citenamefont {Geng}\ \emph {et~al.}(2019)\citenamefont {Geng},
  \citenamefont {Wu}, \citenamefont {Marqu\'es},\ and\ \citenamefont
  {Ackland}}]{Ackland-etal-PhysRevB.100.134109-2019}%
  \BibitemOpen
  \bibfield  {author} {\bibinfo {author} {\bibfnamefont {H.~Y.}\ \bibnamefont
  {Geng}}, \bibinfo {author} {\bibfnamefont {Q.}~\bibnamefont {Wu}}, \bibinfo
  {author} {\bibfnamefont {M.}~\bibnamefont {Marqu\'es}},\ and\ \bibinfo
  {author} {\bibfnamefont {G.~J.}\ \bibnamefont {Ackland}},\ }\bibfield
  {title} {\bibinfo {title} {Thermodynamic anomalies and three distinct
  liquid-liquid transitions in warm dense liquid hydrogen},\ }\href
  {https://doi.org/10.1103/PhysRevB.100.134109} {\bibfield  {journal} {\bibinfo
   {journal} {Phys. Rev. B}\ }\textbf {\bibinfo {volume} {100}},\ \bibinfo
  {pages} {134109} (\bibinfo {year} {2019})}\BibitemShut {NoStop}%
\bibitem [{\citenamefont {Delaney}\ \emph {et~al.}(2006)\citenamefont
  {Delaney}, \citenamefont {Pierleoni},\ and\ \citenamefont
  {Ceperley}}]{DelaneyPierleoniCeperley-PhysRevLett.97.235702-2006}%
  \BibitemOpen
  \bibfield  {author} {\bibinfo {author} {\bibfnamefont {K.~T.}\ \bibnamefont
  {Delaney}}, \bibinfo {author} {\bibfnamefont {C.}~\bibnamefont {Pierleoni}},\
  and\ \bibinfo {author} {\bibfnamefont {D.~M.}\ \bibnamefont {Ceperley}},\
  }\bibfield  {title} {\bibinfo {title} {Quantum monte carlo simulation of the
  high-pressure molecular-atomic crossover in fluid hydrogen},\ }\href
  {https://doi.org/10.1103/PhysRevLett.97.235702} {\bibfield  {journal}
  {\bibinfo  {journal} {Phys. Rev. Lett.}\ }\textbf {\bibinfo {volume} {97}},\
  \bibinfo {pages} {235702} (\bibinfo {year} {2006})}\BibitemShut {NoStop}%
\bibitem [{\citenamefont {Tubman}\ \emph {et~al.}(2015)\citenamefont {Tubman},
  \citenamefont {Liberatore}, \citenamefont {Pierleoni}, \citenamefont
  {Holzmann},\ and\ \citenamefont
  {Ceperley}}]{Tubman-etal-PhysRevLett.115.045301-2015}%
  \BibitemOpen
  \bibfield  {author} {\bibinfo {author} {\bibfnamefont {N.~M.}\ \bibnamefont
  {Tubman}}, \bibinfo {author} {\bibfnamefont {E.}~\bibnamefont {Liberatore}},
  \bibinfo {author} {\bibfnamefont {C.}~\bibnamefont {Pierleoni}}, \bibinfo
  {author} {\bibfnamefont {M.}~\bibnamefont {Holzmann}},\ and\ \bibinfo
  {author} {\bibfnamefont {D.~M.}\ \bibnamefont {Ceperley}},\ }\bibfield
  {title} {\bibinfo {title} {Molecular-atomic transition along the deuterium
  hugoniot curve with coupled electron-ion monte carlo simulations},\ }\href
  {https://doi.org/10.1103/PhysRevLett.115.045301} {\bibfield  {journal}
  {\bibinfo  {journal} {Phys. Rev. Lett.}\ }\textbf {\bibinfo {volume} {115}},\
  \bibinfo {pages} {045301} (\bibinfo {year} {2015})}\BibitemShut {NoStop}%
\bibitem [{\citenamefont {Mazzola}\ and\ \citenamefont
  {Sorella}(2015)}]{MazzolaSorella-PhysRevLett.114.105701-2015}%
  \BibitemOpen
  \bibfield  {author} {\bibinfo {author} {\bibfnamefont {G.}~\bibnamefont
  {Mazzola}}\ and\ \bibinfo {author} {\bibfnamefont {S.}~\bibnamefont
  {Sorella}},\ }\bibfield  {title} {\bibinfo {title} {Distinct metallization
  and atomization transitions in dense liquid hydrogen},\ }\href
  {https://doi.org/10.1103/PhysRevLett.114.105701} {\bibfield  {journal}
  {\bibinfo  {journal} {Phys. Rev. Lett.}\ }\textbf {\bibinfo {volume} {114}},\
  \bibinfo {pages} {105701} (\bibinfo {year} {2015})}\BibitemShut {NoStop}%
\bibitem [{\citenamefont {Mazzola}\ \emph {et~al.}(2018)\citenamefont
  {Mazzola}, \citenamefont {Helled},\ and\ \citenamefont
  {Sorella}}]{Mazzola-etal-PhysRevLett.120.025701-2018}%
  \BibitemOpen
  \bibfield  {author} {\bibinfo {author} {\bibfnamefont {G.}~\bibnamefont
  {Mazzola}}, \bibinfo {author} {\bibfnamefont {R.}~\bibnamefont {Helled}},\
  and\ \bibinfo {author} {\bibfnamefont {S.}~\bibnamefont {Sorella}},\
  }\bibfield  {title} {\bibinfo {title} {Phase diagram of hydrogen and a
  hydrogen-helium mixture at planetary conditions by quantum monte carlo
  simulations},\ }\href {https://doi.org/10.1103/PhysRevLett.120.025701}
  {\bibfield  {journal} {\bibinfo  {journal} {Phys. Rev. Lett.}\ }\textbf
  {\bibinfo {volume} {120}},\ \bibinfo {pages} {025701} (\bibinfo {year}
  {2018})}\BibitemShut {NoStop}%
\bibitem [{\citenamefont {Rillo}\ \emph {et~al.}(2019)\citenamefont {Rillo},
  \citenamefont {Morales}, \citenamefont {Ceperley},\ and\ \citenamefont
  {Pierleoni}}]{Rillo9770}%
  \BibitemOpen
  \bibfield  {author} {\bibinfo {author} {\bibfnamefont {G.}~\bibnamefont
  {Rillo}}, \bibinfo {author} {\bibfnamefont {M.~A.}\ \bibnamefont {Morales}},
  \bibinfo {author} {\bibfnamefont {D.~M.}\ \bibnamefont {Ceperley}},\ and\
  \bibinfo {author} {\bibfnamefont {C.}~\bibnamefont {Pierleoni}},\ }\bibfield
  {title} {\bibinfo {title} {Optical properties of high-pressure fluid hydrogen
  across molecular dissociation},\ }\href
  {https://doi.org/10.1073/pnas.1818897116} {\bibfield  {journal} {\bibinfo
  {journal} {Proceedings of the National Academy of Sciences}\ }\textbf
  {\bibinfo {volume} {116}},\ \bibinfo {pages} {9770} (\bibinfo {year}
  {2019})}\BibitemShut {NoStop}%
\bibitem [{\citenamefont {Karasiev}\ \emph {et~al.}(2018)\citenamefont
  {Karasiev}, \citenamefont {Dufty},\ and\ \citenamefont
  {Trickey}}]{KarasievDuftyTrickey-PhysRevLett.120.076401-2018}%
  \BibitemOpen
  \bibfield  {author} {\bibinfo {author} {\bibfnamefont {V.~V.}\ \bibnamefont
  {Karasiev}}, \bibinfo {author} {\bibfnamefont {J.~W.}\ \bibnamefont
  {Dufty}},\ and\ \bibinfo {author} {\bibfnamefont {S.~B.}\ \bibnamefont
  {Trickey}},\ }\bibfield  {title} {\bibinfo {title} {Nonempirical semilocal
  free-energy density functional for matter under extreme conditions},\ }\href
  {https://doi.org/10.1103/PhysRevLett.120.076401} {\bibfield  {journal}
  {\bibinfo  {journal} {Phys. Rev. Lett.}\ }\textbf {\bibinfo {volume} {120}},\
  \bibinfo {pages} {076401} (\bibinfo {year} {2018})}\BibitemShut {NoStop}%
\bibitem [{\citenamefont {Mermin}(1965)}]{Mermin-PhysRev.137.A1441-1965}%
  \BibitemOpen
  \bibfield  {author} {\bibinfo {author} {\bibfnamefont {N.~D.}\ \bibnamefont
  {Mermin}},\ }\bibfield  {title} {\bibinfo {title} {Thermal properties of the
  inhomogeneous electron gas},\ }\href
  {https://doi.org/10.1103/PhysRev.137.A1441} {\bibfield  {journal} {\bibinfo
  {journal} {Phys. Rev.}\ }\textbf {\bibinfo {volume} {137}},\ \bibinfo {pages}
  {A1441} (\bibinfo {year} {1965})}\BibitemShut {NoStop}%
\bibitem [{\citenamefont {Lankin}\ and\ \citenamefont
  {Norman}(2009)}]{LankinNorman-CPP-2009}%
  \BibitemOpen
  \bibfield  {author} {\bibinfo {author} {\bibfnamefont {A.}~\bibnamefont
  {Lankin}}\ and\ \bibinfo {author} {\bibfnamefont {G.}~\bibnamefont
  {Norman}},\ }\bibfield  {title} {\bibinfo {title} {Density and nonideality
  effects in plasmas},\ }\href {https://doi.org/10.1002/ctpp.200910084}
  {\bibfield  {journal} {\bibinfo  {journal} {Contributions to Plasma Physics}\
  }\textbf {\bibinfo {volume} {49}},\ \bibinfo {pages} {723} (\bibinfo {year}
  {2009})}\BibitemShut {NoStop}%
\bibitem [{\citenamefont {Ma}\ \emph {et~al.}(2019)\citenamefont {Ma},
  \citenamefont {Dai}, \citenamefont {Kang}, \citenamefont {Murillo},
  \citenamefont {Hou}, \citenamefont {Zhao},\ and\ \citenamefont
  {Yuan}}]{eff2019-Murillo}%
  \BibitemOpen
  \bibfield  {author} {\bibinfo {author} {\bibfnamefont {Q.}~\bibnamefont
  {Ma}}, \bibinfo {author} {\bibfnamefont {J.}~\bibnamefont {Dai}}, \bibinfo
  {author} {\bibfnamefont {D.}~\bibnamefont {Kang}}, \bibinfo {author}
  {\bibfnamefont {M.~S.}\ \bibnamefont {Murillo}}, \bibinfo {author}
  {\bibfnamefont {Y.}~\bibnamefont {Hou}}, \bibinfo {author} {\bibfnamefont
  {Z.}~\bibnamefont {Zhao}},\ and\ \bibinfo {author} {\bibfnamefont
  {J.}~\bibnamefont {Yuan}},\ }\bibfield  {title} {\bibinfo {title} {Extremely
  low electron-ion temperature relaxation rates in warm dense hydrogen:
  Interplay between quantum electrons and coupled ions},\ }\href
  {https://doi.org/10.1103/PhysRevLett.122.015001} {\bibfield  {journal}
  {\bibinfo  {journal} {Phys. Rev. Lett.}\ }\textbf {\bibinfo {volume} {122}},\
  \bibinfo {pages} {015001} (\bibinfo {year} {2019})}\BibitemShut {NoStop}%
\bibitem [{\citenamefont {Su}\ and\ \citenamefont {Goddard}(2007)}]{su2007}%
  \BibitemOpen
  \bibfield  {author} {\bibinfo {author} {\bibfnamefont {J.~T.}\ \bibnamefont
  {Su}}\ and\ \bibinfo {author} {\bibfnamefont {W.~A.}\ \bibnamefont
  {Goddard}},\ }\bibfield  {title} {\bibinfo {title} {Excited electron dynamics
  modeling of warm dense matter},\ }\href
  {https://doi.org/10.1103/PhysRevLett.99.185003} {\bibfield  {journal}
  {\bibinfo  {journal} {Phys. Rev. Lett.}\ }\textbf {\bibinfo {volume} {99}},\
  \bibinfo {pages} {185003} (\bibinfo {year} {2007})}\BibitemShut {NoStop}%
\bibitem [{\citenamefont {Frank}\ \emph {et~al.}(1998)\citenamefont {Frank},
  \citenamefont {Hutter}, \citenamefont {Marx},\ and\ \citenamefont
  {Parrinello}}]{roks1998}%
  \BibitemOpen
  \bibfield  {author} {\bibinfo {author} {\bibfnamefont {I.}~\bibnamefont
  {Frank}}, \bibinfo {author} {\bibfnamefont {J.}~\bibnamefont {Hutter}},
  \bibinfo {author} {\bibfnamefont {D.}~\bibnamefont {Marx}},\ and\ \bibinfo
  {author} {\bibfnamefont {M.}~\bibnamefont {Parrinello}},\ }\bibfield  {title}
  {\bibinfo {title} {Molecular dynamics in low-spin excited states},\ }\href
  {https://doi.org/10.1063/1.475804} {\bibfield  {journal} {\bibinfo  {journal}
  {The Journal of Chemical Physics}\ }\textbf {\bibinfo {volume} {108}},\
  \bibinfo {pages} {4060} (\bibinfo {year} {1998})}\BibitemShut {NoStop}%
\bibitem [{\citenamefont {Doltsinis}\ and\ \citenamefont
  {Marx}(2002{\natexlab{a}})}]{cpmd2002}%
  \BibitemOpen
  \bibfield  {author} {\bibinfo {author} {\bibfnamefont {N.~L.}\ \bibnamefont
  {Doltsinis}}\ and\ \bibinfo {author} {\bibfnamefont {D.}~\bibnamefont
  {Marx}},\ }\bibfield  {title} {\bibinfo {title} {Nonadiabatic
  {Car-Parrinello} molecular dynamics},\ }\href
  {https://doi.org/10.1103/PhysRevLett.88.166402} {\bibfield  {journal}
  {\bibinfo  {journal} {Phys. Rev. Lett.}\ }\textbf {\bibinfo {volume} {88}},\
  \bibinfo {pages} {166402} (\bibinfo {year} {2002}{\natexlab{a}})}\BibitemShut
  {NoStop}%
\bibitem [{\citenamefont {Jakob}\ \emph {et~al.}(2009)\citenamefont {Jakob},
  \citenamefont {Reinhard}, \citenamefont {Toepffer},\ and\ \citenamefont
  {Zwicknagel}}]{Jakob-etal-WPMD_2009}%
  \BibitemOpen
  \bibfield  {author} {\bibinfo {author} {\bibfnamefont {B.}~\bibnamefont
  {Jakob}}, \bibinfo {author} {\bibfnamefont {P.-G.}\ \bibnamefont {Reinhard}},
  \bibinfo {author} {\bibfnamefont {C.}~\bibnamefont {Toepffer}},\ and\
  \bibinfo {author} {\bibfnamefont {G.}~\bibnamefont {Zwicknagel}},\ }\bibfield
   {title} {\bibinfo {title} {Wave packet simulations for the
  insulator{\textendash}metal transition in dense hydrogen},\ }\href
  {https://doi.org/10.1088/1751-8113/42/21/214055} {\bibfield  {journal}
  {\bibinfo  {journal} {Journal of Physics A: Mathematical and Theoretical}\
  }\textbf {\bibinfo {volume} {42}},\ \bibinfo {pages} {214055} (\bibinfo
  {year} {2009})}\BibitemShut {NoStop}%
\bibitem [{\citenamefont {Lavrinenko}\ \emph {et~al.}(2019)\citenamefont
  {Lavrinenko}, \citenamefont {Morozov},\ and\ \citenamefont
  {Valuev}}]{Lavrinenko-etal-CPP2019}%
  \BibitemOpen
  \bibfield  {author} {\bibinfo {author} {\bibfnamefont {Y.~S.}\ \bibnamefont
  {Lavrinenko}}, \bibinfo {author} {\bibfnamefont {I.~V.}\ \bibnamefont
  {Morozov}},\ and\ \bibinfo {author} {\bibfnamefont {I.~A.}\ \bibnamefont
  {Valuev}},\ }\bibfield  {title} {\bibinfo {title} {Wave packet molecular
  dynamics–density functional theory method for non-ideal plasma and warm
  dense matter simulations},\ }\href@noop {} {\bibfield  {journal} {\bibinfo
  {journal} {Contributions to Plasma Physics}\ }\textbf {\bibinfo {volume}
  {59}},\ \bibinfo {pages} {e201800179} (\bibinfo {year} {2019})}\BibitemShut
  {NoStop}%
\bibitem [{\citenamefont {Plimpton}(1995)}]{LAMMPS-1995}%
  \BibitemOpen
  \bibfield  {author} {\bibinfo {author} {\bibfnamefont {S.}~\bibnamefont
  {Plimpton}},\ }\bibfield  {title} {\bibinfo {title} {Fast parallel algorithms
  for short-range molecular dynamics},\ }\href
  {https://doi.org/https://doi.org/10.1006/jcph.1995.1039} {\bibfield
  {journal} {\bibinfo  {journal} {Journal of Computational Physics}\ }\textbf
  {\bibinfo {volume} {117}},\ \bibinfo {pages} {1 } (\bibinfo {year}
  {1995})}\BibitemShut {NoStop}%
\bibitem [{Sup()}]{SupportingMaterial}%
  \BibitemOpen
  \href@noop {} {}\bibinfo {note} {See Supplemental Material for the eFF
  results on different densities and heating rates, for the eFF calculations of
  the isotopic effect, for the distribution of S1 lifetimes and for the shapes
  of the electronic orbitals that correspond to exciton states.}\BibitemShut
  {Stop}%
\bibitem [{\citenamefont {Holst}\ \emph {et~al.}(2008)\citenamefont {Holst},
  \citenamefont {Redmer},\ and\ \citenamefont
  {Desjarlais}}]{HolstRedmerDesjarlais-PhysRevB.77.184201-2008}%
  \BibitemOpen
  \bibfield  {author} {\bibinfo {author} {\bibfnamefont {B.}~\bibnamefont
  {Holst}}, \bibinfo {author} {\bibfnamefont {R.}~\bibnamefont {Redmer}},\ and\
  \bibinfo {author} {\bibfnamefont {M.~P.}\ \bibnamefont {Desjarlais}},\
  }\bibfield  {title} {\bibinfo {title} {Thermophysical properties of warm
  dense hydrogen using quantum molecular dynamics simulations},\ }\href
  {https://doi.org/10.1103/PhysRevB.77.184201} {\bibfield  {journal} {\bibinfo
  {journal} {Phys. Rev. B}\ }\textbf {\bibinfo {volume} {77}},\ \bibinfo
  {pages} {184201} (\bibinfo {year} {2008})}\BibitemShut {NoStop}%
\bibitem [{CPM()}]{CPMD-v3.17}%
  \BibitemOpen
  \href@noop {} {}\bibinfo {note} {{CPMD}, version 3.17,
  http://www.cpmd.org.}\BibitemShut {Stop}%
\bibitem [{\citenamefont {Kresse}\ and\ \citenamefont
  {Hafner}(1993)}]{KresseHafner1993}%
  \BibitemOpen
  \bibfield  {author} {\bibinfo {author} {\bibfnamefont {G.}~\bibnamefont
  {Kresse}}\ and\ \bibinfo {author} {\bibfnamefont {J.}~\bibnamefont
  {Hafner}},\ }\bibfield  {title} {\bibinfo {title} {Ab initio molecular
  dynamics for liquid metals},\ }\href
  {https://doi.org/10.1103/PhysRevB.47.558} {\bibfield  {journal} {\bibinfo
  {journal} {Phys. Rev. B}\ }\textbf {\bibinfo {volume} {47}},\ \bibinfo
  {pages} {558} (\bibinfo {year} {1993})}\BibitemShut {NoStop}%
\bibitem [{\citenamefont {Kresse}\ and\ \citenamefont
  {Hafner}(1994)}]{KresseHafner1994}%
  \BibitemOpen
  \bibfield  {author} {\bibinfo {author} {\bibfnamefont {G.}~\bibnamefont
  {Kresse}}\ and\ \bibinfo {author} {\bibfnamefont {J.}~\bibnamefont
  {Hafner}},\ }\bibfield  {title} {\bibinfo {title} {Ab initio
  molecular-dynamics simulation of the liquid-metal-amorphous-semiconductor
  transition in germanium},\ }\href {https://doi.org/10.1103/PhysRevB.49.14251}
  {\bibfield  {journal} {\bibinfo  {journal} {Phys. Rev. B}\ }\textbf {\bibinfo
  {volume} {49}},\ \bibinfo {pages} {14251} (\bibinfo {year}
  {1994})}\BibitemShut {NoStop}%
\bibitem [{\citenamefont {Kresse}\ and\ \citenamefont
  {Furthmuller}(1996)}]{KresseFurthmuller1996}%
  \BibitemOpen
  \bibfield  {author} {\bibinfo {author} {\bibfnamefont {G.}~\bibnamefont
  {Kresse}}\ and\ \bibinfo {author} {\bibfnamefont {J.}~\bibnamefont
  {Furthmuller}},\ }\bibfield  {title} {\bibinfo {title} {Efficiency of
  ab-initio total energy calculations for metals and semiconductors using a
  plane-wave basis set},\ }\href
  {https://doi.org/http://dx.doi.org/10.1016/0927-0256(96)00008-0} {\bibfield
  {journal} {\bibinfo  {journal} {Computational Materials Science}\ }\textbf
  {\bibinfo {volume} {6}},\ \bibinfo {pages} {15 } (\bibinfo {year}
  {1996})}\BibitemShut {NoStop}%
\bibitem [{\citenamefont {Kresse}\ and\ \citenamefont
  {Furthm\"uller}(1996)}]{KresseFurthmuller1999}%
  \BibitemOpen
  \bibfield  {author} {\bibinfo {author} {\bibfnamefont {G.}~\bibnamefont
  {Kresse}}\ and\ \bibinfo {author} {\bibfnamefont {J.}~\bibnamefont
  {Furthm\"uller}},\ }\bibfield  {title} {\bibinfo {title} {Efficient iterative
  schemes for ab initio total-energy calculations using a plane-wave basis
  set},\ }\href {https://doi.org/10.1103/PhysRevB.54.11169} {\bibfield
  {journal} {\bibinfo  {journal} {Phys. Rev. B}\ }\textbf {\bibinfo {volume}
  {54}},\ \bibinfo {pages} {11169} (\bibinfo {year} {1996})}\BibitemShut
  {NoStop}%
\bibitem [{\citenamefont {Peterson}\ \emph {et~al.}(1993)\citenamefont
  {Peterson}, \citenamefont {Schwartz},\ and\ \citenamefont
  {Harris}}]{peterson1993dynamics}%
  \BibitemOpen
  \bibfield  {author} {\bibinfo {author} {\bibfnamefont {E.~S.}\ \bibnamefont
  {Peterson}}, \bibinfo {author} {\bibfnamefont {B.~J.}\ \bibnamefont
  {Schwartz}},\ and\ \bibinfo {author} {\bibfnamefont {C.~B.}\ \bibnamefont
  {Harris}},\ }\bibfield  {title} {\bibinfo {title} {The dynamics of exciton
  tunneling and trapping in condensed xenon on ultrafast time scales},\ }\href
  {https://doi.org/10.1063/1.465286} {\bibfield  {journal} {\bibinfo  {journal}
  {The Journal of Chemical Physics}\ }\textbf {\bibinfo {volume} {99}},\
  \bibinfo {pages} {1693} (\bibinfo {year} {1993})}\BibitemShut {NoStop}%
\bibitem [{\citenamefont {Astashkevich}\ and\ \citenamefont
  {Lavrov}(2015)}]{AstashkevichLavrov-ExcitedH2-lifetimes-2015}%
  \BibitemOpen
  \bibfield  {author} {\bibinfo {author} {\bibfnamefont {S.~A.}\ \bibnamefont
  {Astashkevich}}\ and\ \bibinfo {author} {\bibfnamefont {B.~P.}\ \bibnamefont
  {Lavrov}},\ }\bibfield  {title} {\bibinfo {title} {Lifetimes of
  vibro-rotational levels in excited electronic states of diatomic hydrogen
  isotopologues},\ }\href {https://doi.org/10.1063/1.4921434} {\bibfield
  {journal} {\bibinfo  {journal} {Journal of Physical and Chemical Reference
  Data}\ }\textbf {\bibinfo {volume} {44}},\ \bibinfo {pages} {023105}
  (\bibinfo {year} {2015})}\BibitemShut {NoStop}%
\bibitem [{\citenamefont {Su}\ and\ \citenamefont
  {Goddard}(2009)}]{Su-Goddard-AugerPNAS-2009}%
  \BibitemOpen
  \bibfield  {author} {\bibinfo {author} {\bibfnamefont {J.~T.}\ \bibnamefont
  {Su}}\ and\ \bibinfo {author} {\bibfnamefont {W.~A.}\ \bibnamefont
  {Goddard}},\ }\bibfield  {title} {\bibinfo {title} {Mechanisms of
  {Auger}-induced chemistry derived from wave packet dynamics},\ }\href
  {https://doi.org/10.1073/pnas.0812087106} {\bibfield  {journal} {\bibinfo
  {journal} {Proceedings of the National Academy of Sciences}\ }\textbf
  {\bibinfo {volume} {106}},\ \bibinfo {pages} {1001} (\bibinfo {year}
  {2009})}\BibitemShut {NoStop}%
\bibitem [{\citenamefont {Doltsinis}\ and\ \citenamefont
  {Marx}(2002{\natexlab{b}})}]{doltsinis2002}%
  \BibitemOpen
  \bibfield  {author} {\bibinfo {author} {\bibfnamefont {N.~L.}\ \bibnamefont
  {Doltsinis}}\ and\ \bibinfo {author} {\bibfnamefont {D.}~\bibnamefont
  {Marx}},\ }\bibfield  {title} {\bibinfo {title} {First principles molecular
  dynamics involving excited states and nonadiabatic transitions},\ }\href
  {https://doi.org/10.1142/S0219633602000257} {\bibfield  {journal} {\bibinfo
  {journal} {Journal of Theoretical and Computational Chemistry}\ }\textbf
  {\bibinfo {volume} {01}},\ \bibinfo {pages} {319} (\bibinfo {year}
  {2002}{\natexlab{b}})}\BibitemShut {NoStop}%
\bibitem [{\citenamefont {Jiang}\ \emph {et~al.}(2018)\citenamefont {Jiang},
  \citenamefont {Holtgrewe}, \citenamefont {Geballe}, \citenamefont {Lobanov},
  \citenamefont {Mahmood}, \citenamefont {McWilliams},\ and\ \citenamefont
  {Goncharov}}]{jiang2018insulatormetal}%
  \BibitemOpen
  \bibfield  {author} {\bibinfo {author} {\bibfnamefont {S.}~\bibnamefont
  {Jiang}}, \bibinfo {author} {\bibfnamefont {N.}~\bibnamefont {Holtgrewe}},
  \bibinfo {author} {\bibfnamefont {Z.~M.}\ \bibnamefont {Geballe}}, \bibinfo
  {author} {\bibfnamefont {S.~S.}\ \bibnamefont {Lobanov}}, \bibinfo {author}
  {\bibfnamefont {M.~F.}\ \bibnamefont {Mahmood}}, \bibinfo {author}
  {\bibfnamefont {R.~S.}\ \bibnamefont {McWilliams}},\ and\ \bibinfo {author}
  {\bibfnamefont {A.~F.}\ \bibnamefont {Goncharov}},\ }\href@noop {} {\bibinfo
  {title} {Insulator-metal transition in liquid hydrogen and deuterium}}
  (\bibinfo {year} {2018}),\ \Eprint {https://arxiv.org/abs/1810.01360}
  {arXiv:1810.01360 [cond-mat.mtrl-sci]} \BibitemShut {NoStop}%
\bibitem [{\citenamefont {Shushkov}\ \emph {et~al.}(2012)\citenamefont
  {Shushkov}, \citenamefont {Li},\ and\ \citenamefont
  {Tully}}]{Shushkov-etal-JCP-2012}%
  \BibitemOpen
  \bibfield  {author} {\bibinfo {author} {\bibfnamefont {P.}~\bibnamefont
  {Shushkov}}, \bibinfo {author} {\bibfnamefont {R.}~\bibnamefont {Li}},\ and\
  \bibinfo {author} {\bibfnamefont {J.~C.}\ \bibnamefont {Tully}},\ }\bibfield
  {title} {\bibinfo {title} {Ring polymer molecular dynamics with surface
  hopping},\ }\href@noop {} {\bibfield  {journal} {\bibinfo  {journal} {The
  Journal of Chemical Physics}\ }\textbf {\bibinfo {volume} {137}},\ \bibinfo
  {pages} {22A549} (\bibinfo {year} {2012})}\BibitemShut {NoStop}%
\bibitem [{\citenamefont {Markland}\ and\ \citenamefont
  {Ceriotti}(2018)}]{markland2018nuclear}%
  \BibitemOpen
  \bibfield  {author} {\bibinfo {author} {\bibfnamefont {T.~E.}\ \bibnamefont
  {Markland}}\ and\ \bibinfo {author} {\bibfnamefont {M.}~\bibnamefont
  {Ceriotti}},\ }\bibfield  {title} {\bibinfo {title} {Nuclear quantum effects
  enter the mainstream},\ }\href@noop {} {\bibfield  {journal} {\bibinfo
  {journal} {Nature Reviews Chemistry}\ }\textbf {\bibinfo {volume} {2}},\
  \bibinfo {pages} {1} (\bibinfo {year} {2018})}\BibitemShut {NoStop}%
\bibitem [{\citenamefont {Holmes}\ \emph {et~al.}(1995)\citenamefont {Holmes},
  \citenamefont {Ross},\ and\ \citenamefont
  {Nellis}}]{Holmes-etal-PhysRevB.52.15835-1995}%
  \BibitemOpen
  \bibfield  {author} {\bibinfo {author} {\bibfnamefont {N.~C.}\ \bibnamefont
  {Holmes}}, \bibinfo {author} {\bibfnamefont {M.}~\bibnamefont {Ross}},\ and\
  \bibinfo {author} {\bibfnamefont {W.~J.}\ \bibnamefont {Nellis}},\ }\bibfield
   {title} {\bibinfo {title} {Temperature measurements and dissociation of
  shock-compressed liquid deuterium and hydrogen},\ }\href
  {https://doi.org/10.1103/PhysRevB.52.15835} {\bibfield  {journal} {\bibinfo
  {journal} {Phys. Rev. B}\ }\textbf {\bibinfo {volume} {52}},\ \bibinfo
  {pages} {15835} (\bibinfo {year} {1995})}\BibitemShut {NoStop}%
\bibitem [{\citenamefont {Houtput}\ \emph {et~al.}(2019)\citenamefont
  {Houtput}, \citenamefont {Tempere},\ and\ \citenamefont
  {Silvera}}]{houtput2019finite_PRB}%
  \BibitemOpen
  \bibfield  {author} {\bibinfo {author} {\bibfnamefont {M.}~\bibnamefont
  {Houtput}}, \bibinfo {author} {\bibfnamefont {J.}~\bibnamefont {Tempere}},\
  and\ \bibinfo {author} {\bibfnamefont {I.~F.}\ \bibnamefont {Silvera}},\
  }\bibfield  {title} {\bibinfo {title} {Finite-element simulation of the
  liquid-liquid transition to metallic hydrogen},\ }\href
  {https://doi.org/10.1103/PhysRevB.100.134106} {\bibfield  {journal} {\bibinfo
   {journal} {Phys. Rev. B}\ }\textbf {\bibinfo {volume} {100}},\ \bibinfo
  {pages} {134106} (\bibinfo {year} {2019})}\BibitemShut {NoStop}%
\bibitem [{\citenamefont {Morales}\ \emph {et~al.}(2010)\citenamefont
  {Morales}, \citenamefont {Pierleoni}, \citenamefont {Schwegler},\ and\
  \citenamefont {Ceperley}}]{Morales-etal-PNAS-2010}%
  \BibitemOpen
  \bibfield  {author} {\bibinfo {author} {\bibfnamefont {M.~A.}\ \bibnamefont
  {Morales}}, \bibinfo {author} {\bibfnamefont {C.}~\bibnamefont {Pierleoni}},
  \bibinfo {author} {\bibfnamefont {E.}~\bibnamefont {Schwegler}},\ and\
  \bibinfo {author} {\bibfnamefont {D.~M.}\ \bibnamefont {Ceperley}},\
  }\bibfield  {title} {\bibinfo {title} {Evidence for a first-order
  liquid-liquid transition in high-pressure hydrogen from ab initio
  simulations},\ }\href {https://doi.org/10.1073/pnas.1007309107} {\bibfield
  {journal} {\bibinfo  {journal} {Proceedings of the National Academy of
  Sciences}\ }\textbf {\bibinfo {volume} {107}},\ \bibinfo {pages} {12799}
  (\bibinfo {year} {2010})}\BibitemShut {NoStop}%
\bibitem [{\citenamefont {Zaghoo}(2018)}]{Zaghoo-PhysRevE.97.043205-2018}%
  \BibitemOpen
  \bibfield  {author} {\bibinfo {author} {\bibfnamefont {M.}~\bibnamefont
  {Zaghoo}},\ }\bibfield  {title} {\bibinfo {title} {Dynamic conductivity and
  partial ionization in dense fluid hydrogen},\ }\href
  {https://doi.org/10.1103/PhysRevE.97.043205} {\bibfield  {journal} {\bibinfo
  {journal} {Phys. Rev. E}\ }\textbf {\bibinfo {volume} {97}},\ \bibinfo
  {pages} {043205} (\bibinfo {year} {2018})}\BibitemShut {NoStop}%
\bibitem [{\citenamefont {Norman}\ \emph {et~al.}(2019)\citenamefont {Norman},
  \citenamefont {Saitov},\ and\ \citenamefont
  {Sartan}}]{NormanSaitovSartan-CPP-2019}%
  \BibitemOpen
  \bibfield  {author} {\bibinfo {author} {\bibfnamefont {G.~E.}\ \bibnamefont
  {Norman}}, \bibinfo {author} {\bibfnamefont {I.~M.}\ \bibnamefont {Saitov}},\
  and\ \bibinfo {author} {\bibfnamefont {R.~A.}\ \bibnamefont {Sartan}},\
  }\bibfield  {title} {\bibinfo {title} {Metastable molecular fluid hydrogen at
  high pressures},\ }\href {https://doi.org/10.1002/ctpp.201800173} {\bibfield
  {journal} {\bibinfo  {journal} {Contributions to Plasma Physics}\ }\textbf
  {\bibinfo {volume} {59}},\ \bibinfo {pages} {e201800173} (\bibinfo {year}
  {2019})}\BibitemShut {NoStop}%
\bibitem [{\citenamefont {Stegailov}\ and\ \citenamefont
  {Zhilyaev}(2016)}]{StegailovZhilyaev-MolPhys-2016}%
  \BibitemOpen
  \bibfield  {author} {\bibinfo {author} {\bibfnamefont {V.~V.}\ \bibnamefont
  {Stegailov}}\ and\ \bibinfo {author} {\bibfnamefont {P.~A.}\ \bibnamefont
  {Zhilyaev}},\ }\bibfield  {title} {\bibinfo {title} {Warm dense gold:
  effective ion–ion interaction and ionisation},\ }\href
  {https://doi.org/10.1080/00268976.2015.1105390} {\bibfield  {journal}
  {\bibinfo  {journal} {Molecular Physics}\ }\textbf {\bibinfo {volume}
  {114}},\ \bibinfo {pages} {509} (\bibinfo {year} {2016})}\BibitemShut
  {NoStop}%
\bibitem [{\citenamefont {Norman}\ \emph {et~al.}(2015)\citenamefont {Norman},
  \citenamefont {Saitov},\ and\ \citenamefont
  {Stegailov}}]{NormanSaitovStegailov-PPT-CPP-2015}%
  \BibitemOpen
  \bibfield  {author} {\bibinfo {author} {\bibfnamefont {G.~E.}\ \bibnamefont
  {Norman}}, \bibinfo {author} {\bibfnamefont {I.~M.}\ \bibnamefont {Saitov}},\
  and\ \bibinfo {author} {\bibfnamefont {V.~V.}\ \bibnamefont {Stegailov}},\
  }\bibfield  {title} {\bibinfo {title} {Plasma-plasma and liquid-liquid
  first-order phase transitions},\ }\href
  {https://doi.org/10.1002/ctpp.201400088} {\bibfield  {journal} {\bibinfo
  {journal} {Contributions to Plasma Physics}\ }\textbf {\bibinfo {volume}
  {55}},\ \bibinfo {pages} {215} (\bibinfo {year} {2015})}\BibitemShut
  {NoStop}%
\bibitem [{\citenamefont {Norman}\ and\ \citenamefont
  {Saitov}(2019)}]{NormanSaitov-CPP-PPT50-2019}%
  \BibitemOpen
  \bibfield  {author} {\bibinfo {author} {\bibfnamefont {G.~E.}\ \bibnamefont
  {Norman}}\ and\ \bibinfo {author} {\bibfnamefont {I.~M.}\ \bibnamefont
  {Saitov}},\ }\bibfield  {title} {\bibinfo {title} {Plasma phase transition
  (by the fiftieth anniversary of the prediction)},\ }\href
  {https://doi.org/10.1002/ctpp.201800182} {\bibfield  {journal} {\bibinfo
  {journal} {Contributions to Plasma Physics}\ }\textbf {\bibinfo {volume}
  {59}},\ \bibinfo {pages} {e201800182} (\bibinfo {year} {2019})}\BibitemShut
  {NoStop}%
\bibitem [{\citenamefont {Pickard}\ and\ \citenamefont
  {Needs}(2007)}]{pickard2007structure}%
  \BibitemOpen
  \bibfield  {author} {\bibinfo {author} {\bibfnamefont {C.~J.}\ \bibnamefont
  {Pickard}}\ and\ \bibinfo {author} {\bibfnamefont {R.~J.}\ \bibnamefont
  {Needs}},\ }\bibfield  {title} {\bibinfo {title} {Structure of phase {III} of
  solid hydrogen},\ }\href@noop {} {\bibfield  {journal} {\bibinfo  {journal}
  {Nature Physics}\ }\textbf {\bibinfo {volume} {3}},\ \bibinfo {pages} {473}
  (\bibinfo {year} {2007})}\BibitemShut {NoStop}%
\bibitem [{\citenamefont {McMahon}\ and\ \citenamefont
  {Ceperley}(2011)}]{McMahonCeperly-PhysRevLett.106.165302-2011}%
  \BibitemOpen
  \bibfield  {author} {\bibinfo {author} {\bibfnamefont {J.~M.}\ \bibnamefont
  {McMahon}}\ and\ \bibinfo {author} {\bibfnamefont {D.~M.}\ \bibnamefont
  {Ceperley}},\ }\bibfield  {title} {\bibinfo {title} {Ground-state structures
  of atomic metallic hydrogen},\ }\href
  {https://doi.org/10.1103/PhysRevLett.106.165302} {\bibfield  {journal}
  {\bibinfo  {journal} {Phys. Rev. Lett.}\ }\textbf {\bibinfo {volume} {106}},\
  \bibinfo {pages} {165302} (\bibinfo {year} {2011})}\BibitemShut {NoStop}%
\bibitem [{\citenamefont {Dias}\ and\ \citenamefont
  {Silvera}(2017)}]{DiasSilvera-Science-2017}%
  \BibitemOpen
  \bibfield  {author} {\bibinfo {author} {\bibfnamefont {R.~P.}\ \bibnamefont
  {Dias}}\ and\ \bibinfo {author} {\bibfnamefont {I.~F.}\ \bibnamefont
  {Silvera}},\ }\bibfield  {title} {\bibinfo {title} {Observation of the
  {Wigner-Huntington} transition to metallic hydrogen},\ }\href
  {https://doi.org/10.1126/science.aal1579} {\bibfield  {journal} {\bibinfo
  {journal} {Science}\ }\textbf {\bibinfo {volume} {355}},\ \bibinfo {pages}
  {715} (\bibinfo {year} {2017})}\BibitemShut {NoStop}%
\bibitem [{\citenamefont {Saitov}(2019)}]{Saitov-JETPLett-2019}%
  \BibitemOpen
  \bibfield  {author} {\bibinfo {author} {\bibfnamefont {I.~M.}\ \bibnamefont
  {Saitov}},\ }\bibfield  {title} {\bibinfo {title} {Metastable conducting
  crystalline hydrogen at high pressure},\ }\href
  {https://doi.org/10.1134/S0021364019150116} {\bibfield  {journal} {\bibinfo
  {journal} {JETP Letters}\ }\textbf {\bibinfo {volume} {110}},\ \bibinfo
  {pages} {206} (\bibinfo {year} {2019})}\BibitemShut {NoStop}%
\bibitem [{\citenamefont {Rohlfing}\ and\ \citenamefont
  {Louie}(2000)}]{rohlfing2000electron}%
  \BibitemOpen
  \bibfield  {author} {\bibinfo {author} {\bibfnamefont {M.}~\bibnamefont
  {Rohlfing}}\ and\ \bibinfo {author} {\bibfnamefont {S.~G.}\ \bibnamefont
  {Louie}},\ }\bibfield  {title} {\bibinfo {title} {Electron-hole excitations
  and optical spectra from first principles},\ }\href@noop {} {\bibfield
  {journal} {\bibinfo  {journal} {Physical Review B}\ }\textbf {\bibinfo
  {volume} {62}},\ \bibinfo {pages} {4927} (\bibinfo {year}
  {2000})}\BibitemShut {NoStop}%
\end{thebibliography}

%

\end{document}


\title{Supplementary materials for: ``Non-adiabatic effects and exciton-like states during insulator-to-metal transition in warm dense hydrogen''}

\author{I.~D.~Fedorov}
    \email[]{fedorov.id@mipt.ru}
    \affiliation{Joint Institute for High Temperatures of the Russian Academy of Sciences (JIHT RAS), Izhorskaya 13 Building 2, Moscow 125412, Russian Federation}
    \affiliation{Moscow Institute of Physics and Technologies National Research University (MIPT NRU), Institutskij pereulok 9, Dolgoprudny Moscow region 141700, Russian Federation}
\author{N.~D.~Orekhov}
    \affiliation{Joint Institute for High Temperatures of the Russian Academy of Sciences (JIHT RAS), Izhorskaya 13 Building 2, Moscow 125412, Russian Federation}
    \affiliation{Moscow Institute of Physics and Technologies National Research University (MIPT NRU), Institutskij pereulok 9, Dolgoprudny Moscow region 141700, Russian Federation}
\author{V.~V.~Stegailov}
    \email[]{stegailov.vv@mipt.ru}
    \affiliation{Joint Institute for High Temperatures of the Russian Academy of Sciences (JIHT RAS), Izhorskaya 13 Building 2, Moscow 125412, Russian Federation}
    \affiliation{Moscow Institute of Physics and Technologies National Research University (MIPT NRU), Institutskij pereulok 9, Dolgoprudny Moscow region 141700, Russian Federation}
    \affiliation{National Research University Higher School of Economics (NRU HSE), Myasnitskaya ulitsa 20, Moscow 101000 Russian Federation}

\date{\today} 
\maketitle 

\begin{figure}
    \centering
    \includegraphics[width=\textwidth]{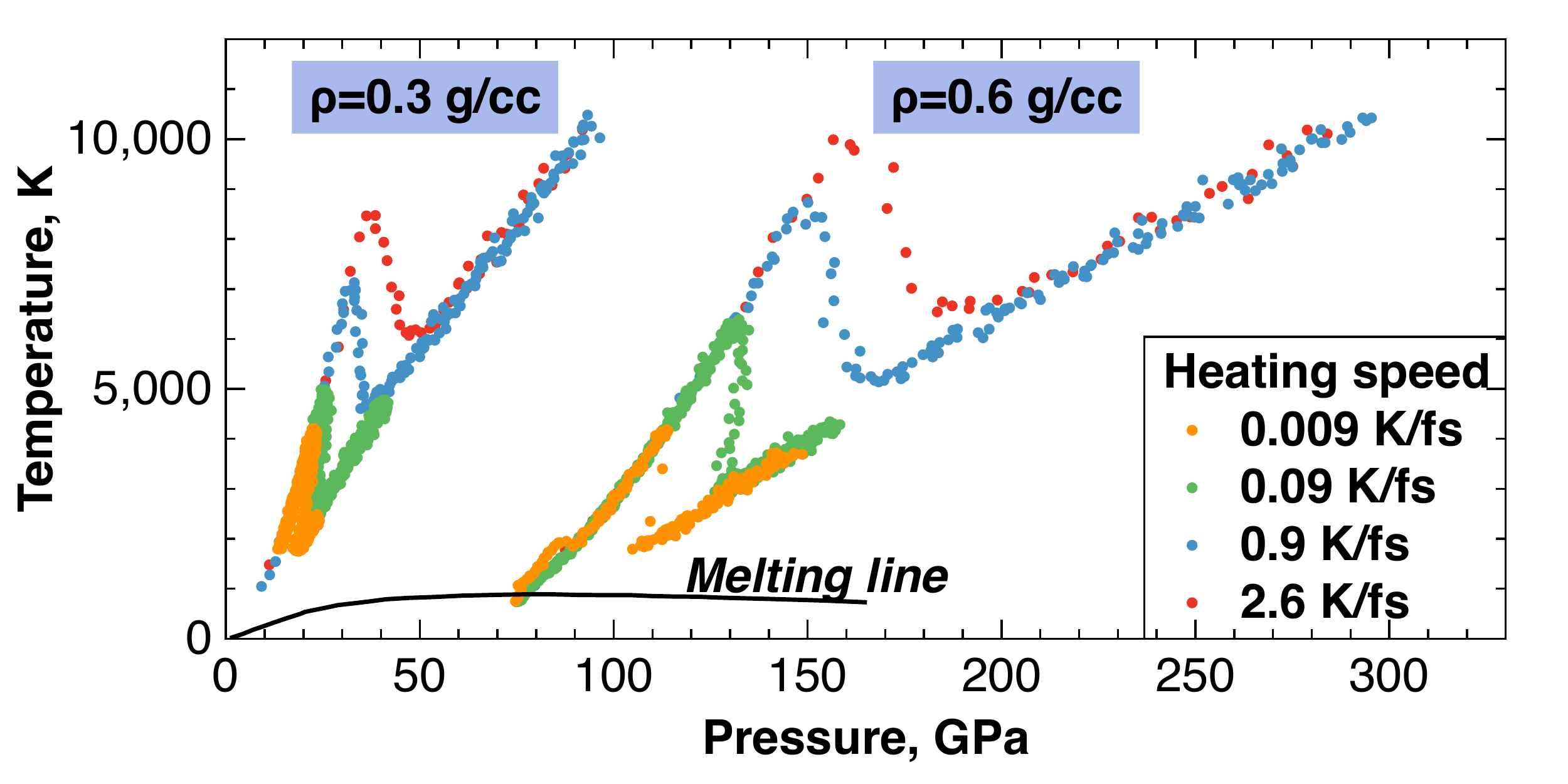}
    \caption{The temperature of nuclei along the isochores during heating of the fluid H$_2$ eFF model. The melting line of solid hydrogen is shown for the reference.}
    \label{fig:eFF_PT_common}
\end{figure}

\section{Isohoric heating of fluid H$_2$ in the eFF model at different rates}

To calculate the instantaneous temperatures of nuclei ($T_n$) and electrons ($T_e$) in the eFF model, we consider electron and nuclei degrees of freedom separately that results in the following expressions for the total kinetic energies of nuclear and electron subsystems:

\begin{equation}
K_{n}=\sum \frac{1}{2}m_{n}\dot{{\bf r}}^{2}_{n},
\label{eq_Kn}
\end{equation}

\begin{equation}
K_{e}=\sum \frac{1}{2}m_{e}\dot{{\bf r}}^{2}_{e}+\sum \frac{1}{2}\frac{3}{4}m_{e}\dot{s}^{2}_{e},
\label{eq_Ke}
\end{equation}
where $s_e$ corresponds to the size of the electron wave packet, $\dot{{\bf r}}_n$ and $\dot{{\bf r}}_e$ are the translational velocities of nuclei and electrons. The instantaneous temperatures are calculated as the measures of the average kinetic energy of motion:

\begin{equation}
T_{n}=\frac{2}{3k_{B}N_{n}}\left \langle K_{n} \right \rangle,~~~
T_{e}=\frac{2}{4k_{B}N_{e}}\left \langle K_{e} \right \rangle.
\label{eq_T}
\end{equation}

The isochoric heating of the system of 2000 nuclei and 2000 electrons is performed by rescaling the velocities of the nuclei every 50 MD steps. The timestep is 0.005~fs.

Fig.~\ref{fig:eFF_PT_common} shows the isochores of fluid H$_2$ in the eFF model. Isochoric heating starts from the states with the molecules in the ground state. Different heating rates are considered. 

\begin{figure}
\centering
\includegraphics[width=\textwidth]{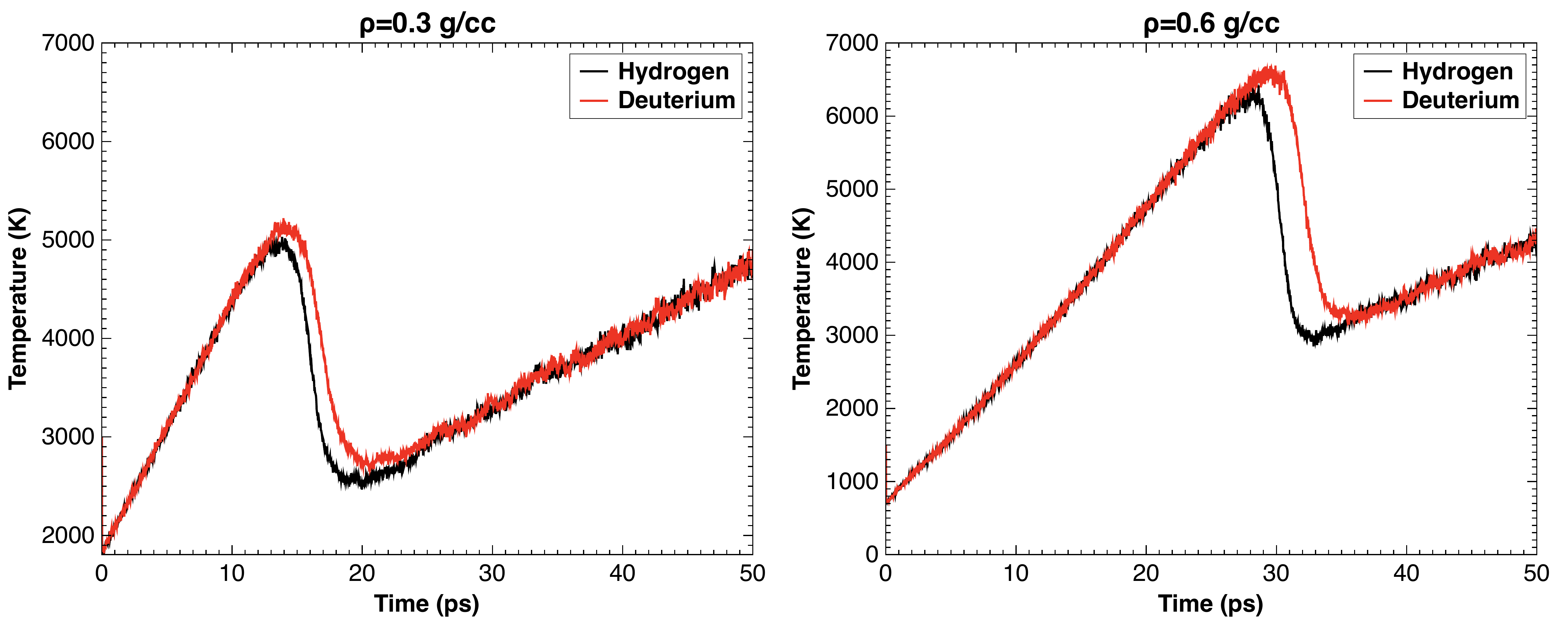}
\caption{The temperature of nuclei during isochoric heating of the fluid H$_2$/D$_2$ eFF model for the H$_2$ density 0.3~g/cc and 0.6~g/cc (the corresponding D$_2$ densities are twice larger).}
\label{fig:isotopic_effect}
\end{figure}

The first kink at each isochore corresponds to the beginning of the vibronic excitation of electrons. These points are the beginning of IMT at each of the isochores. The second kink at each isochore corresponds to the end of the energy transfer from nuclei to electrons when their temperatures become equal $T_n = T_e$. The temperature of IMT (the first kink) decreases for lower heating rates.

Here we would like to emphasize that in the eFF model after the vibronic excitation of the first molecule all the molecules in the system become excited very rapidly (in the ``avalanche'' mode). Presumably, it should be assumed to be an artifact of eFF since the rate of the spontaneous de-excitation (the radiationless non-adiabatic internal conversion) in eFF seems to be zero.

\section{Isotopic effect in eFF model}

Fig.~\ref{fig:isotopic_effect} shows the results of calculations that illustrate the isotopic effect in the eFF model. At the H$_2$ density of $\rho=0.6$~g/cc (and the twice larger D$_2$ density) the isotopic effect is about 400~K that is close to the experimental observations \href{https://doi.org/10.1103/PhysRevB.98.104102}{[Zaghoo, Husband, and Silvera, Phys. Rev. B 98, 104102 (2018)]}.

\begin{figure}
\centering
\includegraphics[width=0.8\textwidth]{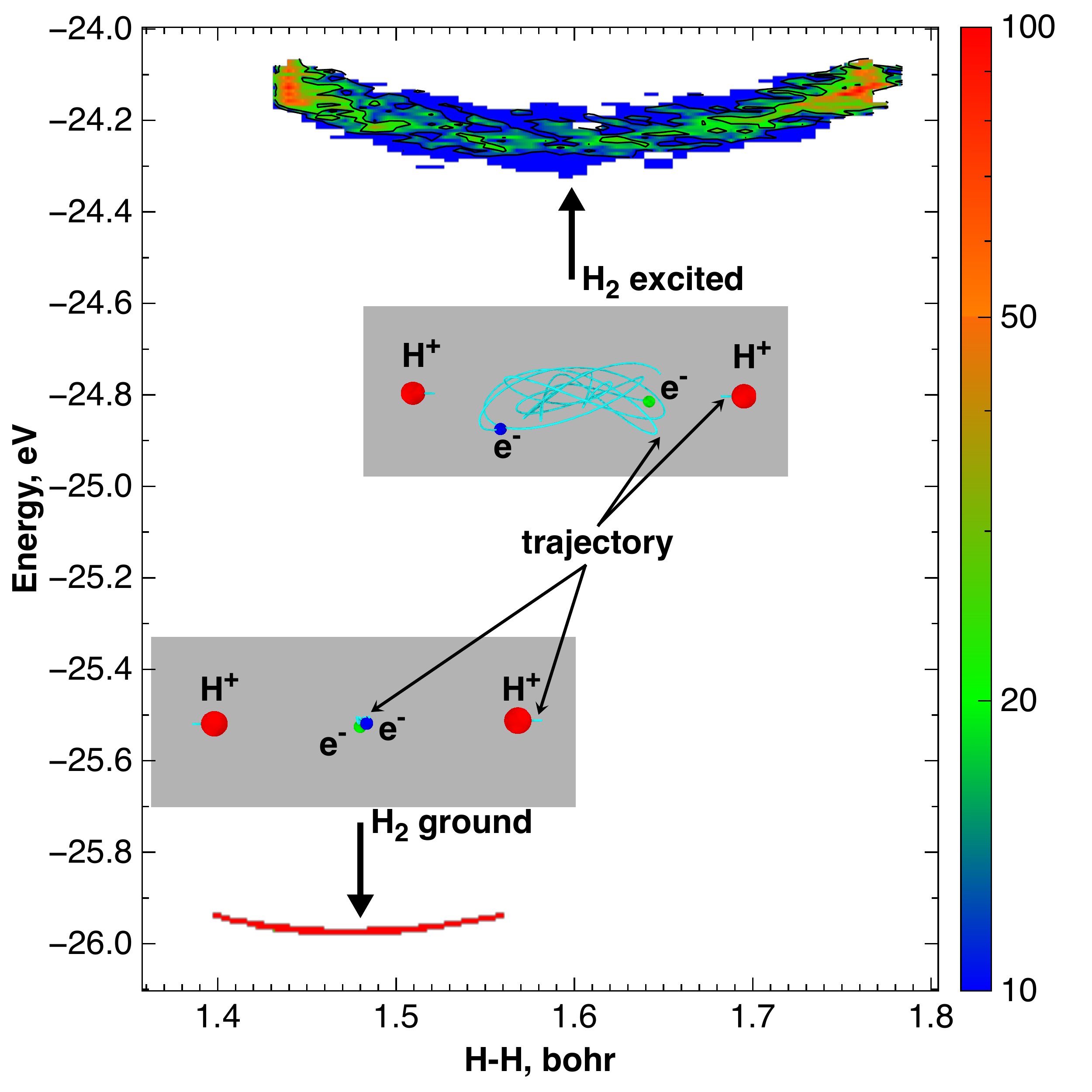}
\caption{The dependence of the potential energy of an isolated H$_2$ molecule during oscillations in the ground state and in the excited state. Two sets of color points show the probability distributions for finding a molecule with a certain value of its potential energy at a certain H-H distance (the color corresponds to the count values, see the colormap). The lower distribution shows a molecule in the ground state when the electron kinetic energy is close to zero. The upper distribution shows a molecule in the excited state with the effective temperature of electrons is 3000~K. Two insets show the corresponding trajectory fragments (250~fs each) for the nuclei and for the centres of electron wave packets.}
\label{fig:h2_term}
\end{figure}

\section{The ground state and the excited state of a hydrogen molecule in the eFF model}

In order to clarify the meaning of the excited states of H$_2$ molecules in the eFF model, we put on the plot (Fig.~\ref{fig:h2_term}) the values of the potential energy of the molecule and the corresponding values of H-H distances for the ground state case and for the excited case. Electrons, being significantly lighter than nuclei, move much faster than nuclei and create some averaged field that forms an effective potential between two ions. This analysis shows two distinct states of the isolated H$_2$ molecule. It is interesting that in the excited state we see a clear minimum that justifies the stability of the molecule. 

\begin{figure}
\centering
\includegraphics[width=0.7\textwidth]{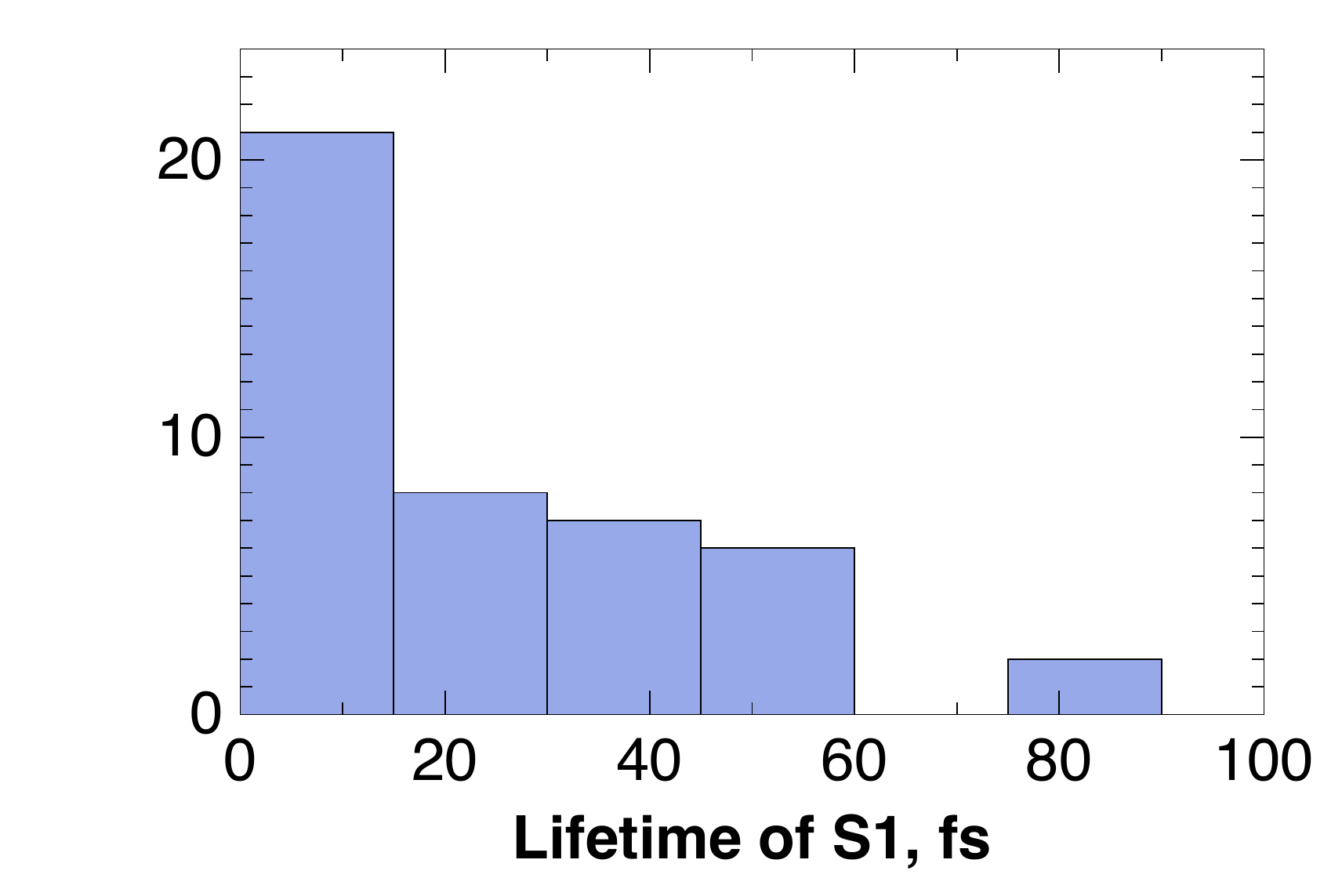}
\caption{The S1 lifetimes distribution from 10 MD trajectories for the system of 30 H$_2$ molecules at $T=1300$~K and $\rho$=0.6~g/cc.}
\label{fig:lifetimedist}
\end{figure}

\section{Vibronic excitation of the first molecule in the system}

An animation that illustrates the vibronic excitation of the first molecule in the system at $\rho=0.3$~g/cc during isochoric heating  presented in the file \verb|h2_excitation.mp4|. The large green balls represent the hydrogen nuclei. The centers of the electronic wave-packets are shown as the small balls that are green at low kinetic energy. They become red when the kinetic energy of the corresponding electron becomes higher than the threshold value.

\section{Distribution of S1 lifetimes during surface hopping MD}

10 different equilibrated atomic configurations with 60 atoms have been obtained for 1300~K and 0.6~g/cc. MD with surface hopping is calculated from each of these initial conditions for 500~fs. In this ensemble of MD trajectories we find 44 excitations from S0 to S1 and back to S0. Fig.~\ref{fig:lifetimedist} shows the corresponding distribution of S1-state lifetimes.

\section{Visualization of the excitons revealed via ROKS DFT}

\begin{figure}
\centering
\begin{subfigure}[b]{0.7\textwidth}
\includegraphics[width=0.5\textwidth]{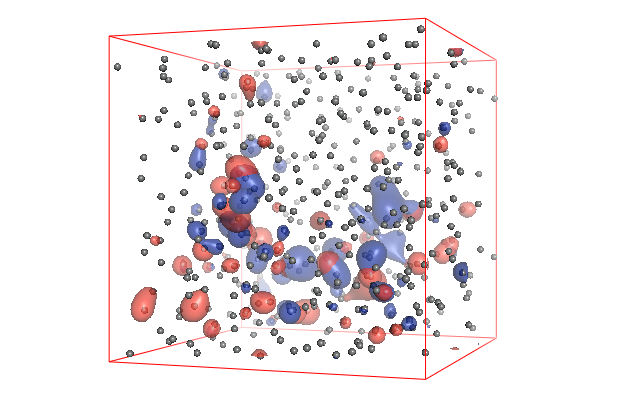}
    \caption{480 nuclei, isolevel 0.0013}
\end{subfigure}
\begin{subfigure}[b]{0.7\textwidth}
    \includegraphics[width=\textwidth]{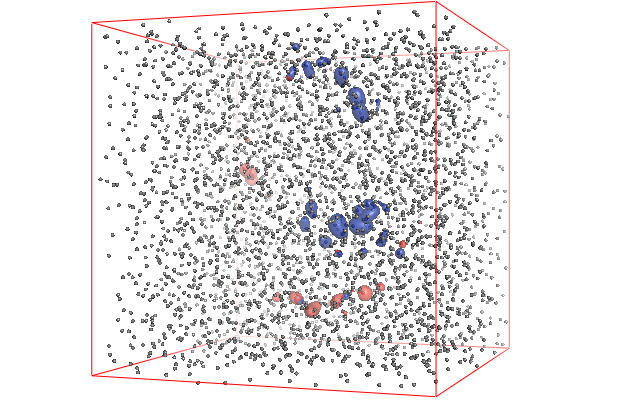}
    \caption{3840 nuclei, isolevel 0.0013}
\end{subfigure}
\begin{subfigure}[b]{0.7\textwidth}
    \includegraphics[width=\textwidth]{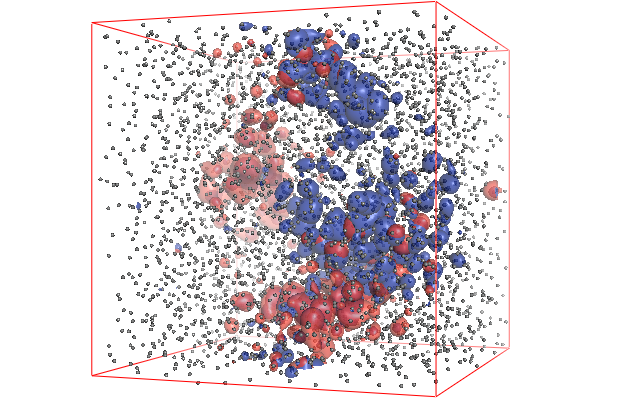}
    \caption{3840 nuclei, isolevel $0001625 = 0.0013 / 8$}
\end{subfigure}
\caption{Isosurfaces of SOMO-1 and SOMO-2 orbitals for $\rho=0.7$ g/cc.}
\label{fig:system-size-07}
\end{figure}
\begin{figure}
\centering
\begin{subfigure}[b]{0.7\textwidth}
\includegraphics[width=0.5\textwidth]{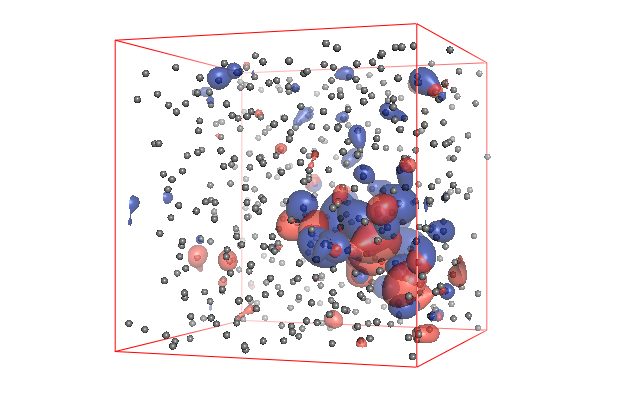}
    \caption{480 nuclei, isolevel 0.0013}
\end{subfigure}
\begin{subfigure}[b]{0.7\textwidth}
    \includegraphics[width=\textwidth]{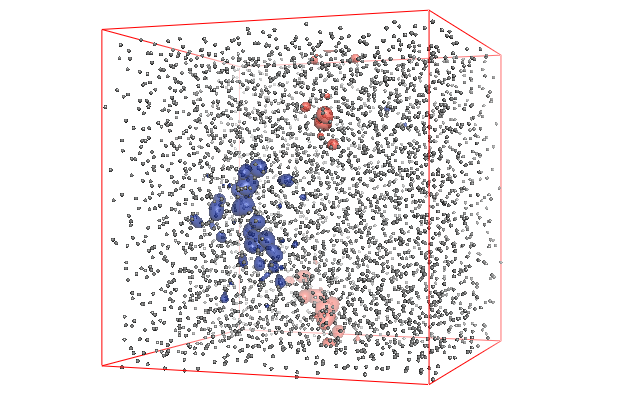}
    \caption{3840 nuclei, isolevel 0.0013}
\end{subfigure}
\begin{subfigure}[b]{0.7\textwidth}
    \includegraphics[width=\textwidth]{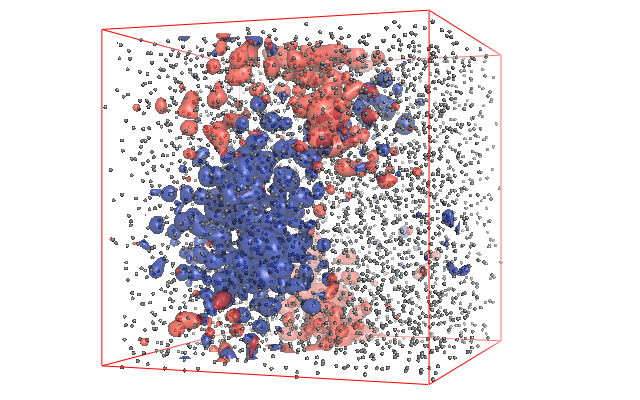}
    \caption{3840 nuclei, isolevel $0001625 = 0.0013 / 8$}
\end{subfigure}
\caption{Isosurfaces of SOMO-1 and SOMO-2 orbitals for $\rho=0.8$ g/cc.}
\label{fig:system-size-08}
\end{figure}
\begin{figure}
\centering
\begin{subfigure}[b]{0.7\textwidth}
\includegraphics[width=0.5\textwidth]{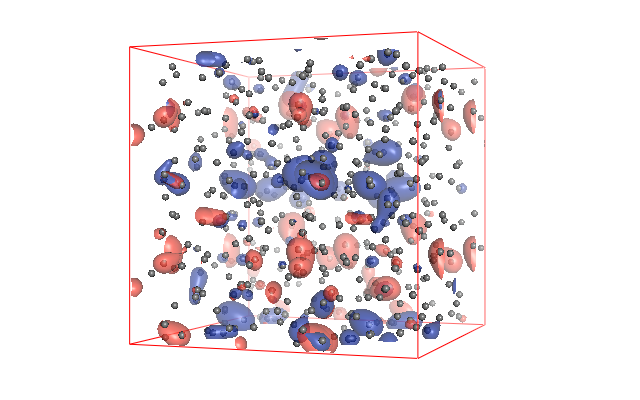}
    \caption{480 nuclei, isolevel 0.0013}
\end{subfigure}
\begin{subfigure}[b]{0.7\textwidth}
    \includegraphics[width=\textwidth]{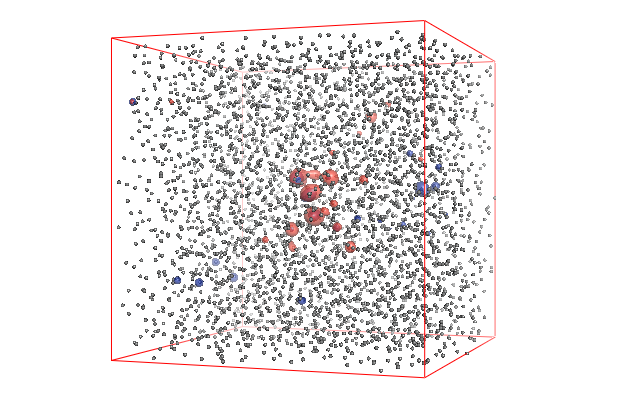}
    \caption{3840 nuclei, isolevel 0.0013}
\end{subfigure}
\begin{subfigure}[b]{0.7\textwidth}
    \includegraphics[width=\textwidth]{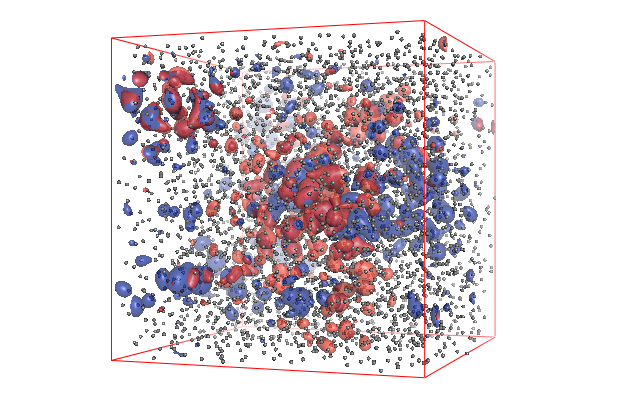}
    \caption{3840 nuclei, isolevel $0001625 = 0.0013 / 8$}
\end{subfigure}
\caption{Isosurfaces of SOMO-1 and SOMO-2 orbitals for $\rho=0.9$ g/cc.}
\label{fig:system-size-09}
\end{figure}
\begin{figure}
\centering
\begin{subfigure}[b]{0.7\textwidth}
\includegraphics[width=0.5\textwidth]{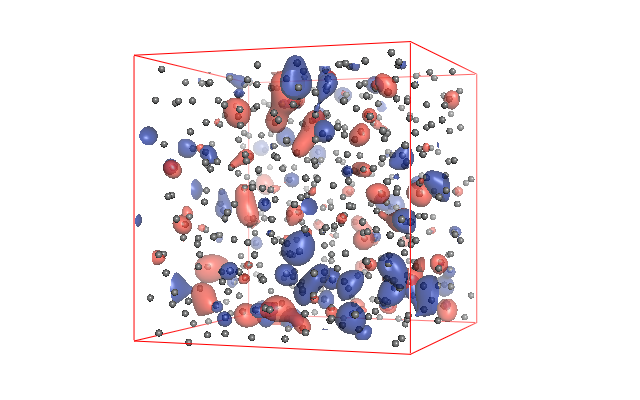}
    \caption{480 nuclei, isolevel 0.0013}
\end{subfigure}
\begin{subfigure}[b]{0.7\textwidth}
    \includegraphics[width=\textwidth]{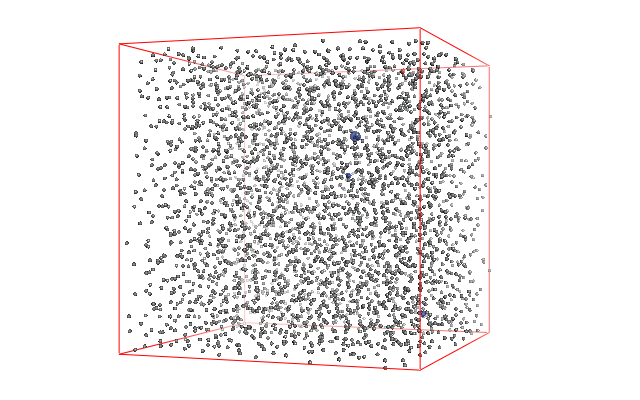}
    \caption{3840 nuclei, isolevel 0.0013}
\end{subfigure}
\begin{subfigure}[b]{0.7\textwidth}
    \includegraphics[width=\textwidth]{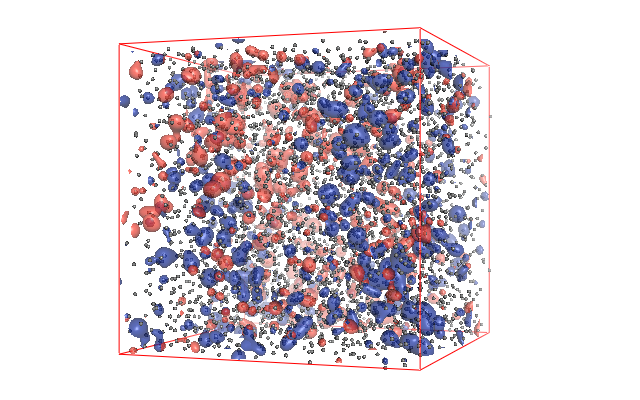}
    \caption{3840 nuclei, isolevel $0001625 = 0.0013 / 8$}
\end{subfigure}
\caption{Isosurfaces of SOMO-1 and SOMO-2 orbitals for $\rho=1.0$ g/cc.}
\label{fig:system-size-10}
\end{figure}

In order to illustrate the size effects on Fig.~\ref{fig:system-size-07}-\ref{fig:system-size-10} we present the comparisons of SOMO-1/2 orbitals at different densities in the systems of 480 atoms and in the systems of 3840 atoms (8 time larger). The systems with different sizes have been equilibrated independently that is why the excitons are different. It is only their visible size that we compare. 

One can see that the excitons at $\rho = 0.7-0.8$~g/cc are localized and there is no evident dependence of their spacial extention on the system size.  At $\rho = 1.0$~g/cc the exciton is delocilized for both system sizes (due to the spreading of the orbitals in the larger system the exciton visualisation needs a lower isolevel value). At $\rho = 0.9$~g/cc we observe a threshold situation between the localized and delocalized cases.

